\documentclass[structabstract]{aa}  
\pdfoutput=1

\usepackage{graphicx}
\usepackage{natbib}
\bibpunct{(}{)}{;}{a}{}{,}
\usepackage{upgreek}
\usepackage{amsmath}
\usepackage{amssymb}
\usepackage{txfonts}
\usepackage[linkcolor=blue, bookmarks=true, bookmarksopen=true,
a4paper=true, citecolor=blue, urlcolor=blue, colorlinks=true,
pagecolor=blue, breaklinks=true, plainpages=false]{hyperref}
\usepackage[all]{hypcap}

\newcommand{\earth}{\mathrm{\oplus}}
\newcommand{\sub}[1]{_\mathrm{#1}}
\newcommand{\dif}{\mathrm{d}}
\mathchardef\mhyphen="2D

\begin{document}

 \title{Near-infrared emission from sublimating dust in collisionally active debris disks}
 \titlerunning{NIR emission from sublimating dust in debris disks}

  \author{R. van Lieshout\inst{1}
	\and
	C. Dominik\inst{1,2}
	\and
	M. Kama\inst{1}
	\and
	M. Min\inst{1}
	}

  \institute{Astronomical Institute ``Anton Pannekoek'', University of Amsterdam,
	    P.O. Box 94249, 1090 GE Amsterdam, The Netherlands\\
	    \email{r.vanlieshout@uva.nl}
	\and
	    Department of Astrophysics/IMAPP, Radboud University Nijmegen, P.O. Box 9010, 6500 GL Nijmegen, The Netherlands
	    }

\abstract
  {
  Hot exozodiacal dust
  is thought to be responsible for
  excess near-infrared (NIR) emission
  emanating from the innermost parts of some debris disks.
  The origin
  of this dust, however,
  is still a matter of debate.
  }
  {
  We test whether
  hot exozodiacal dust
  can be supplied from an exterior parent belt by Poynting--Robertson \mbox{(P--R)} drag,
  paying special attention to
  the pile-up of dust that occurs
  due to the interplay of \mbox{P--R} drag and dust sublimation.
  Specifically, we investigate whether pile-ups still occur when collisions are taken into account,
  and if they can explain
  the observed NIR excess.
  }
  {
  We compute the steady-state distribution of dust
  in the inner disk by solving the continuity equation.
  First, we derive an analytical solution under a number of simplifying assumptions.
  Second, we develop a numerical debris disk model
  that for the first time treats the complex interaction of collisions, \mbox{P--R} drag, and sublimation in a self-consistent way.
  From the resulting dust distributions we generate
  thermal emission spectra
  and compare these to observed excess NIR fluxes.
  }
  {
  We confirm that P--R drag always supplies a small amount of dust to the sublimation zone,
  but find that a fully consistent treatment
  yields a maximum amount of dust that is about 7 times lower than that given by analytical estimates.
  The NIR excess due this material is much smaller
  ($\lesssim$10$^{-3}$ for \mbox{A-type} stars with parent belts at $\gtrsim$1~AU)
  than the values derived from interferometric observations ($\sim$10$^{-2}$).
  Pile-up of dust still occurs when collisions are considered,
  but its effect on the NIR flux is insignificant.
  Finally, the cross-section in the innermost regions is clearly dominated by barely bound grains.
  }
  {}

  \keywords{zodiacal dust -- circumstellar matter}

  \maketitle

\section{Introduction}

Circumstellar dust in debris disks reveals the location and dynamical state of larger bodies,
and hence sheds light on the architecture of planetary systems
in the aftermath of planet formation
(see \citealt{2008ARA&A..46..339W} for a review).
The dust can be studied
by observing its infrared and (sub-)millimeter emission,
as well as the stellar radiation it scatters,
and is usually found at large distances from the star \citep[tens of AUs,][]{2009ApJS..181..197C}.
Recently, interferometric observations
have found excess near-infrared (NIR) emission
emanating from
the innermost parts of several debris disks,
which has been interpreted as thermal emission from hot ($>$1000~K) dust
\citep[see Tbl.~\ref{tbl:obs} for an overview]{2001ApJ...559.1147C,2006A&A...452..237A,
2008A&A...487.1041A,2009ApJ...704..150A,2007A&A...475..243D,2009ApJ...691.1896A,2011A&A...534A...5D,2012A&A...546L...9D}.
This material is known as hot exozodiacal dust.
Its origin,
and hence
what it can tell us about
planet formation,
is still unclear.
In this work, we investigate one possible scenario to explain hot exozodiacal dust.

\begin{table*}[!t]
 \centering
 \caption{NIR interferometric detections of hot exozodiacal dust, together with associated outer debris belt locations}
 \label{tbl:obs}
 \begin{tabular}{cc|cr@{ $\pm$ }lccc|ccc}
 \hline
 Object & Sp. type & Band & \multicolumn{2}{c}{Excess} & FOV\tablefootmark{a} & Instrument & Refs. & Outer belt distance\tablefootmark{b} & Refs. \\
  & & & \multicolumn{2}{c}{[\%]} & [AU] & & & [AU] & \\
 \hline
  Vega & A0V & $H$ & 1.23 & 0.53 & 6 & IOTA/IONIC & D11
  & 10--14, 80 & D00, S13 \\
  Vega & A0V & $K$ & 1.26 & 0.27 & 3 & CHARA/FLUOR & A06, A13
  & 10--14, 80 & D00, S13 \\
  Vega & A0V & $K$ & \multicolumn{2}{c}{$5^{+1}_{-2}\;\;$} & 4 & PTI & C01
  & 10--14, 80 & D00, S13 \\
  $\upzeta$~Aql & A0V & $K$ & 1.69 & 0.27 & 10 & CHARA/FLUOR & A08, A13
  & no detectable outer belt & A08, P09 \\
  $\upbeta$~Leo & A3V & $K$ & 0.94 & 0.26 & 4 & CHARA/FLUOR & Ak09, A13
  & 19 & C06 \\
  $\uplambda$~Gem & A3V & $K$ & 0.74 & 0.17 & 12 & CHARA/FLUOR & A13
  & no detectable outer belt & M09, G13 \\
  Fomalhaut & A4V & $K$ & 0.88 & 0.12 & 6 & VLTI/VINCI & Ab09
  & 2, 8--11, 133 & K05, L13, S13 \\
  $\upbeta$~Pic\tablefootmark{c} & A6V & $H$ & 1.37 & 0.16 & 4 & VLTI/PIONIER & D12
  & 10--40\tablefootmark{d} & L94, P97 \\
  $\upbeta$~Pic\tablefootmark{c} & A6V & $K$ & 0.76 & 0.49 & 1.3 & VLTI/VINCI & D04, D12
  & 10--40\tablefootmark{d} & L94, P97 \\
  $\upalpha$~Aql & A7V & $K$ & 3.07 & 0.24 & 2 & CHARA/FLUOR & A13
  & no detectable outer belt & A13 \\
  $\upalpha$~Cep & A7IV & $K$ & 0.87 & 0.18 & 6 & CHARA/FLUOR & A13
  & no detectable outer belt & C05 \\
  $\upeta$~Lep\tablefootmark{e} & F1V & $K$ & 0.89 & 0.21 & 6 & CHARA/FLUOR & A13
  & 1--16, 18 & L09, E13 \\
  110~Her\tablefootmark{e} & F6V & $K$ & 0.94 & 0.25 & 8 & CHARA/FLUOR & A13
  & 70--500 & M13 \\
  10~Tau\tablefootmark{e} & F9V & $K$ & 1.21 & 0.11 & 6 & CHARA/FLUOR & A13
  & $>$5.8 & T08 \\
  $\upxi$~Boo\tablefootmark{e} & G8V & $K$ & 0.74 & 0.20 & 3 & CHARA/FLUOR & A13
  & no detectable outer belt & A13 \\
  $\uptau$~Cet & G8V & $K$ & 0.98 & 0.18 & 1.5 & CHARA/FLUOR & D07, A13
  & 10--55 & G04 \\
  $\upkappa$~CrB\tablefootmark{e} & K1IV & $K$ & 1.18 & 0.20 & 12 & CHARA/FLUOR & A13
  & 20, 41 & B13 \\
 \hline
 \end{tabular}
 \tablefoot{ \\
 \tablefoottext{a}FOV denotes the approximate linear field-of-view radius at half maximum. \\
 \tablefoottext{b}For the outer belt distance ($r\sub{0}$ in our models)
 we list literature estimates of the radial distance to (the inner edge of) ``cold'' and ``warm'' outer belts,
 derived from SED fitting and/or resolved imaging. \\
 \tablefoottext{c}The NIR excess of $\upbeta$~Pic contains a significant contribution
 from stellar light scattered by the outer belt \citep{2012A&A...546L...9D}. \\
 \tablefoottext{d}The debris disk around $\upbeta$~Pic is seen edge-on, making it hard to determine the parent belt location.
 The values given mark the radial range in which the particle density is derived to decrease. \\
 \tablefoottext{e}For $\upeta$~Lep, 110~Her, 10~Tau, $\upxi$~Boo, and $\upkappa$~CrB,
 the possibility that the observed NIR excess is due to
 a low-mass companion within the field-of-view cannot be excluded \citep{2013A&A...555A.104A}.
 }
 \tablebib{
 (A06)~\citet{2006A&A...452..237A};
 (A08)~\citet{2008A&A...487.1041A};
 (Ab09)~\citet{2009ApJ...704..150A};
 (Ak09)~\citet{2009ApJ...691.1896A};
 (A13)~\citet{2013A&A...555A.104A};
 (B13)~\citet{2013MNRAS.431.3025B};
 (C01)~\citet{2001ApJ...559.1147C};
 (C05)~\citet{2005ApJ...634.1372C};
 (C06)~\citet{2006ApJS..166..351C};
 (D00)~\citet{2000MNRAS.314..702D};
 (D04)~\citet{2004A&A...426..601D};
 (D07)~\citet{2007A&A...475..243D};
 (D11)~\citet{2011A&A...534A...5D};
 (D12)~\citet{2012A&A...546L...9D};
 (E13)~\citet{2013A&A...555A..11E};
 (G04)~\citet{2004MNRAS.351L..54G};
 (G13)~\citet{2013ApJ...768...25G};
 (K05)~\citet{2005Natur.435.1067K};
 (L94)~\citet{1994Natur.369..628L};
 (L09)~\citet{2009ApJ...705...89L};
 (L13)~\citet{2013A&A...555A.146L};
 (M09)~\citet{2009ApJ...699.1067M};
 (M13)~\citet{2013A&A...557A..58M};
 (P97)~\citet{1997A&A...327.1123P};
 (P09)~\citet{2009ApJ...698.1068P};
 (S13)~\citet{2013ApJ...763..118S};
 (T08)~\citet{2008ApJ...674.1086T}.
 }
\end{table*}

Dust grains in debris disks have relatively short lifetimes,
due to their
destruction by collisions and
removal by radiation forces.
The detection of these grains around mature stars
therefore implies the existence of a mechanism that continuously replenishes them.
Cold dust populations at large distances from the star
can be maintained by a collisional cascade grinding down much larger bodies
that act as a reservoir of mass \citep{1993prpl.conf.1253B}.
Closer to the star, however,
the pace at which material is processed by collisions is much higher,
and hence the lifetime of a debris belt in collisional equilibrium is much shorter there
\citep{2003ApJ...598..626D,2007ApJ...658..569W}.
For this reason, hot exozodiacal dust cannot be explained by in-situ planetesimal belts
\citep{2007ApJ...658..569W,2013A&A...555A.146L},\footnote{
\citet{2013MNRAS.433.2334K} find that
``warm'' exozodiacal dust around solar-type stars can be explained by in-situ planetesimal belts.
This type of exozodiacal dust is detected at mid-infrared wavelengths and
has a typical temperature of a few hundred K, placing it around 1~AU from the star.}
and a different mechanism is needed to replenish it,
and/or the lifetime of the dust needs to be extended by some process.

Many of the systems that exhibit NIR excess
also feature a debris belt at a large distance from the star (see Tbl.~\ref{tbl:obs}).
Inward transport of material from an outer belt
may therefore be a natural explanation for
the existence of hot exozodiacal dust.
A possible transportation mechanism is Poynting--Robertson (\mbox{P--R}) drag
\citep[see, e.g.,][]{1979Icar...40....1B}.
Because \mbox{P--R} drag acts on a timescale that is much longer than that of collisions,
it is sometimes disfavored as possible mechanism
for maintaining exozodiacal dust \citep[e.g.,][]{2006A&A...452..237A}.
However, as long as there are no mechanisms that prevent inward migration,
a small amount of dust is always transported to the innermost part of the disk \citep{2005A&A...433.1007W},
where it produces a NIR signal.

Morphological models of exozodiacal dust disks, constrained by the NIR observations,
indicate that the hot dust is concentrated in
a sharply peaked ring,
whose inner boundary is determined by dust sublimation \citep{2011A&A...534A...5D,2013ApJ...763..119M,2013A&A...555A.146L}.
The process of dust sublimation may therefore play an important role in shaping exozodiacal clouds.
\citet{2009Icar..201..395K} find that
the interplay between \mbox{P--R} drag and dust sublimation
can lead to a local enhancement of dust in the sublimation zone,
leading to radial distributions of dust reminiscent of those found by the morphological models.
However, they only investigate this pile-up effect
for drag-dominated systems, in which collisions are unimportant,
and it is unclear what happens to the phenomenon
if collisions are taken into account.

In this work,
we examine whether it is possible to maintain a pile-up of dust in the sublimation zone
of a collisionally active debris disk,
and whether such a pile-up could explain the exozodiacal NIR emission observed very close to some stars.
To do this, we compute
the steady-state distribution of dust in the inner parts of debris disks,
under the influence of collisions, \mbox{P--R} drag, and sublimation,
by solving the continuity equation.
First, we
find an analytical solution, using a number of  simplifying assumptions (Sect.~\ref{s:analytic}).
Subsequently, we solve the continuity equation numerically
using a debris disk model that for the first time treats the complex interaction of
collisions, \mbox{P--R} drag, and sublimation in a self-consistent way (Sect.~\ref{s:numeric}).
From the obtained steady-state dust distributions, we compute emission spectra
to compare with observational data (Sect.~\ref{s:seds}).
We discuss our findings in Sect.~\ref{s:discussion}, and give conclusions in Sect.~\ref{s:conclusions}.
Details of the numerical techniques employed by the debris disk model are given in Appendix~\ref{s:app_num},
model verification tests are described in Appendix~\ref{s:app_verif},
and the post-processing of model output into useful physical quantities in described in Appendix~\ref{s:app_postproc}.

\section{Analytical constraints}
\label{s:analytic}

In this section we analytically investigate the distribution of dust
in the inner regions of debris disks.
We focus on the radial distribution of material,
by assuming that (1)~the disk is axisymmetric,
and (2)~all particles have the same size.
Throughout this work,
radial distributions are expressed
in terms of
vertical geometrical optical depth,
which is defined as the surface density of cross-section.\footnote{
The geometrical optical depth corresponds to the true vertical optical depth only for
an extinction efficiency of unity ($Q\sub{ext} = 1$) for all particle sizes.
}
Under the two assumptions listed above, it is given by
\begin{equation}
  \label{eq:tau_eff_approx}
  \tau\sub{geo}(r)
  = \frac{ \sigma n(r) }{ 2 \pi r }.
\end{equation}
Here, $r$ is the radial distance from the central star,
$\sigma$ is the cross-section of a particle,
and $n(r)$ is the one dimensional number density
(i.e., the particle number density integrated over disk height and azimuth).

We consider three processes affecting the evolution of dust particles in debris disks:
collisions, \mbox{P--R} drag, and sublimation.
The strategy for our analytical estimates is as follows.
First, we review the balance between \mbox{P--R} drag and collisions (without considering sublimation),
and calculate the inward flux of material due to these two effects (Sect.~\ref{s:prcoll}).
Subsequently, we consider the interplay of \mbox{P--R} drag and sublimation (ignoring collisions),
which can lead to the pile-up of dust in the sublimation zone (Sect.~\ref{s:prsubl}).
Finally, we investigate whether collisions can be neglected in the innermost
parts of a debris disk,
and estimate
the radial distribution of dust in a disk in which all three processes operate (Sect.~\ref{s:pilecoll}).
At the end of the section we briefly summarize our findings (Sect.~\ref{s:analitic_sum}).

\subsection{Poynting--Robertson drag and collisions}
\label{s:prcoll}

The balance between \mbox{P--R} drag and collisions was studied analytically by \citet{2005A&A...433.1007W}.
Because of its importance to the present study, we summarize the main arguments of this work here.
The model assumes that
(1)~there is a source of dust at distance $ r\sub{0} $ with a geometrical optical depth of $ \tau\sub{geo}(r\sub{0}) $,
(2)~dust particles follow circular orbits,
(3)~collisions are always destructive,
and (4)~all dust grains have the same size.
These assumptions lead to simple expressions for the 
timescales on which \mbox{P--R} drag and collisions typically act.

The \mbox{P--R} drag timescale $ t\sub{PR} $ is defined as the time it takes for
a particle on a circular orbit to spiral from a given distance $r$ to the central star.
It is given by \citep[e.g.,][]{1979Icar...40....1B}
\begin{equation}
  \label{eq:t_P--R}
  t\sub{PR}(r) = \frac{ c r^2 }{ 4 G M\sub{\star} \beta },
\end{equation}
where $ c $ is the speed of light, $ G $ is the gravitational constant, $ M\sub{\star} $ is the stellar mass,
and $ \beta $ is the ratio of the norms of the direct radiation pressure force and the gravitational force on a particle
($ \beta = | F\sub{rp} / F\sub{g} | $).\footnote{
Assuming circular orbits is valid for particles with low \mbox{$ \beta $-ratios},
while the small particles most relevant to our study have high \mbox{$ \beta $-ratios}.
We relax this simplifying assumption in our numerical model (Sect.~\ref{s:numeric}).}

The collisional timescale $ t\sub{coll} $ indicates the average time between two collisions for a given particle.
\citet{2005A&A...433.1007W} finds
\begin{equation}
  \label{eq:t_coll}
  t\sub{coll}(r) = \frac{ t\sub{orb}(r) }{ 4 \pi \tau\sub{geo}(r) },
\end{equation}
where $ t\sub{orb} $ is the orbital period of a circular orbit, given by
$ t\sub{orb}(r) = 2 \pi \sqrt{ r^3 / ( G M\sub{\star} ) } $.
This equation is valid for particles that are on circular orbits,
and whose most important collisional partners are of similar size.\footnote{
Equation~\ref{eq:t_coll} ignores a factor $ \sqrt{ 1 / (1 - \beta) } $
in the orbital period of radiation pressure affected particles,
but this only changes the collisional timescale
by a factor of
about 0.7 for $ \beta = 0.5 $.
}

\subsubsection{The radial distribution due to \mbox{P--R} drag and collisions}
\label{s:prcoll_rad}

Under the assumptions listed above,
there is an analytical solution to the continuity equation,
balancing the migration of particles due to \mbox{P--R} drag with the destruction of dust by collisions.
\citet{2005A&A...433.1007W} finds that the steady-state solution is
\begin{subequations}
  \label{eq:tau_eff_prcoll}
  \begin{align}
    \tau\sub{geo}(r) &= \frac{ \tau\sub{geo}(r\sub{0}) }{ 1 + 4 \eta\sub{0} ( 1 - \sqrt{ r / r\sub{0} } ) },
    \qquad r \leq  r\sub{0}, \\
    \label{eq:eta0}
    \eta\sub{0} &= \frac{ c \tau\sub{geo}(r\sub{0}) }{ 2 \beta } \sqrt{ \frac{ r\sub{0} }{ G M\sub{\star} } }.
  \end{align}
\end{subequations}
The parameter $ \eta\sub{0} $ characterizes the density of the parent belt.
It is defined such that for $ \eta\sub{0} = 1 $
the collisional and \mbox{P--R} drag timescales
are equal at $ r\sub{0} $.
Disks with $ \eta\sub{0} > 1 $ are collision dominated at $ r\sub{0} $,
while disks with $ \eta\sub{0} < 1 $ are drag dominated at $ r\sub{0} $.
Most
debris disks with observed outer belts have $ \eta\sub{0} > 10 $ \citep{2005A&A...433.1007W}.

\begin{figure}[!t]
  \includegraphics[width=\linewidth]{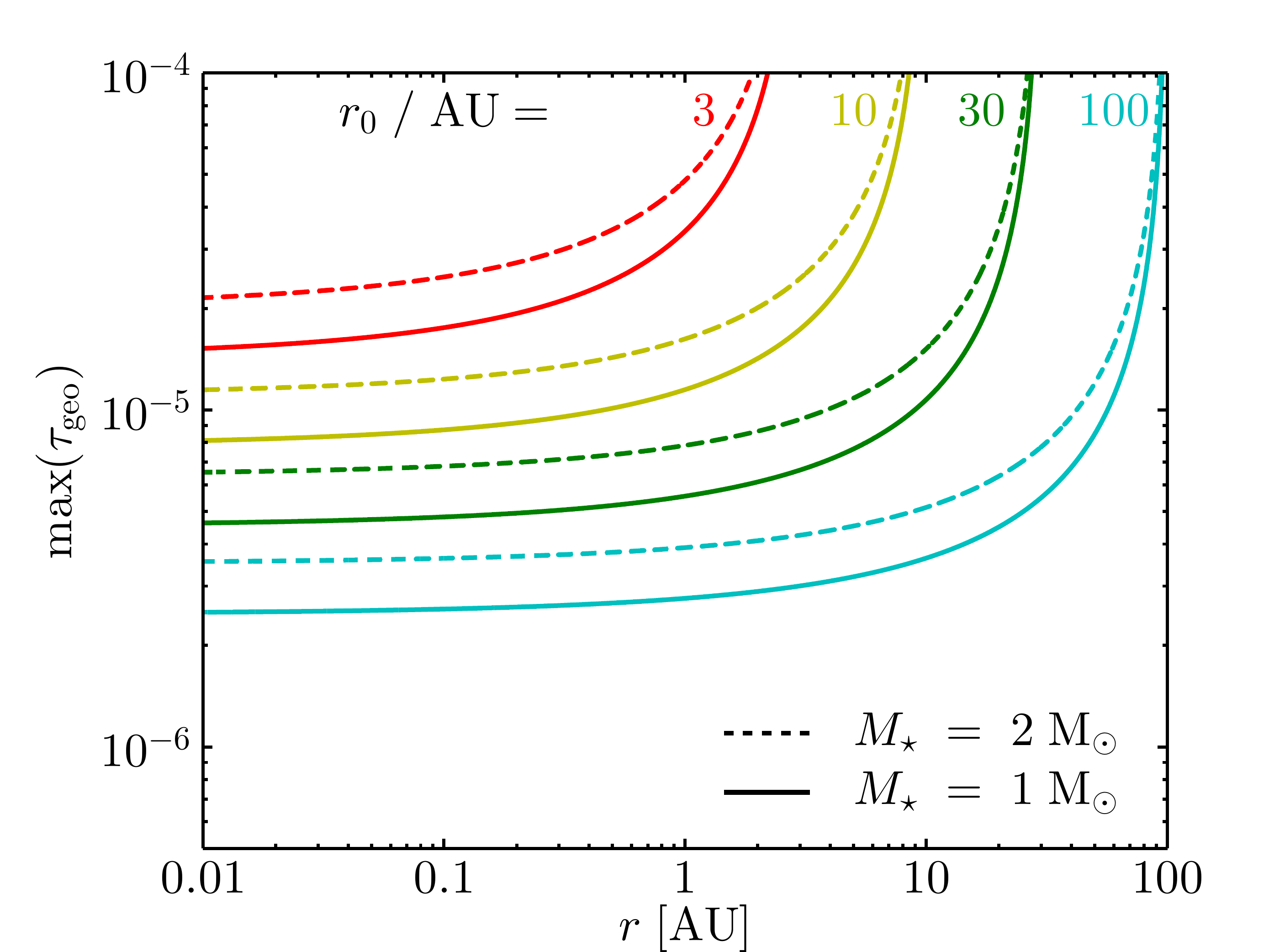}
  \caption{
  The maximum geometrical optical depth as a function of distance to the star (Eq.~\ref{eq:max_tau_eff}).
  The profiles are derived from the analytical model of \citet{2005A&A...433.1007W},
  in which dust is produced by a source at radius $r\sub{0}$
  and subsequently migrates inward due to \mbox{P--R} drag,
  while suffering destruction from mutual collisions.
  The solid lines correspond to solar-mass stars, the dashed lines to 2~M$\sub{\odot}$ stars.
  Profiles for different parent belt locations are shown in different colors.
  All profiles assume dust grains with $ \beta = 0.5 $.
  }
  \label{fig:max_tau}
\end{figure}

The balance between \mbox{P--R} drag and collisions is self-limiting:
a denser parent belt produces more dust drifting inwards,
but this dust also suffers more mutual collisions that eliminate grains on their way in.
This results in a maximum geometrical optical depth profile
for $ \eta\sub{0} \gg 1 $ (i.e., a very dense parent belt)
of
\begin{equation}
  \label{eq:max_tau_eff}
  \max[\tau\sub{geo}(r)]
    = \frac{ \sqrt{ G M\sub{\star} } \, \beta }{ 2 c \left( \sqrt{ r\sub{0} } - \sqrt{ r } \right) },
    \qquad r \leq  r\sub{0}.
\end{equation}
Figure~\ref{fig:max_tau} shows examples of this radial profile
for different parent belt locations and host star masses,
all for dust grains with $ \beta = 0.5 $.
The value for $ \beta $ was set to the blowout limit:
particles with $ \beta > 0.5 $ leave the system on hyperbolic paths
after they are released from a large parent body on a circular orbit.
Therefore, \mbox{P--R} drag is the most efficient for $ \beta = 0.5 $,
and for a given system, this value corresponds to the maximum \mbox{$\tau\sub{geo}$-profile}.\footnote{
Debris disks around stars with a strong stellar wind have higher values of $\tau\sub{geo}$.
We ignore stellar wind in our present analysis, and discuss its effects in Sect.~\ref{s:discuss_wind_drag}.
}

\subsubsection{The inward flux of material}
\label{s:prcoll_flux}

Since dust can pile up
close to the star due to sublimation
(to be discussed in Sec.~\ref{s:pileup}),
$ \tau\sub{geo} $ may exceed the upper limit given by Eq.~\ref{eq:max_tau_eff} in the sublimation zone.
The material that piles up, however, is supplied from further out by \mbox{P--R} drag.
To investigate the properties of the pile-up,
it is therefore useful to assess the inward flux of material.

In the case of a uniform grain size,
the inward particle flux due to \mbox{P--R} drag
(i.e., the number of particles passing through a ring at radius $r$ per unit of time)
can be expressed as
\begin{equation}
  \label{eq:ndot_pr}
  \varphi\sub{PR}(r)
  = - n(r) \dot{r}\sub{PR} ( r ),
\end{equation}
where $\dot{r}\sub{PR}$ is the \mbox{P--R} drag velocity,
counted positively towards $r>0$.
If the particle orbits are circular,
radial migration due to \mbox{P--R} drag
is described by \citep[e.g.,][]{1979Icar...40....1B}
\begin{equation}
  \label{eq:rdot_pr}
  \dot{r}\sub{PR}(r)
  = - \frac{ 2 G M\sub{\star} \beta }{ c r }.
\end{equation}
The grain parameter $ \beta $ can be expressed as \citep[e.g.,][]{1979Icar...40....1B}
\begin{equation}
  \label{eq:beta}
  \beta  = \frac{ L\sub{\star} }{ 4 \pi c G M\sub{\star} }
    \frac{ Q\sub{pr} \sigma }{ m }.
\end{equation}
Here, $L\sub{\star}$ is the stellar luminosity,
$m$ is the mass of an individual dust grain, 
and $Q\sub{pr}$ is the radiation pressure efficiency averaged over the stellar spectrum.
Combining Eqs.~\ref{eq:tau_eff_approx}, \ref{eq:ndot_pr}, \ref{eq:rdot_pr}, and \ref{eq:beta},
and multiplying by particle mass $m$, gives the inward (collision-limited) \mbox{P--R} drag mass flux
\citep[cf.][]{2011ApJ...732L...3R}
\begin{equation}
  \label{eq:mass_flux}
  \dot{ M }\sub{PR}(r)
  = \frac{ L\sub{\star} }{ c^2 } Q\sub{pr} \tau\sub{geo}(r).
\end{equation}

The maximum mass flux can now be found
by substituting Eq.~\ref{eq:max_tau_eff} for $ \tau\sub{geo}(r) $ in Eq.~\ref{eq:mass_flux}, which yields
\begin{equation}
  \label{eq:max_mass_flux}
  \max \left[ \dot{ M }\sub{PR}(r) \right]
  = \frac{ \sqrt{ G M\sub{\star} } \, L\sub{\star} \beta Q\sub{pr} }
    { 2 c^3 \left( \sqrt{ r\sub{0} } - \sqrt{ r } \right) },
  \qquad r \leq  r\sub{0}.
\end{equation}
Figure~\ref{fig:max_tau} shows that $ \max[ \tau\sub{geo}(r) ] $ levels off in the innermost part of the system ($ r \ll r\sub{0} $),
at the radii where exozodiacal dust is
seen.
The mass flux corresponding to this plateau is
\begin{equation}
  \begin{split}
  \label{eq:max_mass_flux_r0}
  \max [ \dot{ M }\sub{PR}( r = 0 ) ]
  & \approx 5.6  \times 10^{-13}
  \, \biggl( \frac{ M\sub{\star} }{ \mathrm{1~M}\sub{\odot} } \biggr)^{1/2}
  \, \biggl( \frac{ L\sub{\star} }{ \mathrm{1~L}\sub{\odot} } \biggr) \\
  & \quad \times
  \biggl( \frac{ r\sub{0} }{ \mathrm{1~AU} } \biggr)^{-1/2}
  \, \biggl( \frac{ Q\sub{pr} }{ 1 } \biggr)
  \, \biggl( \frac{ \beta }{ 0.5 } \biggr)
  \; \mathrm{M}\sub{\earth} \; \mathrm{yr}^{-1}.
  \end{split}
\end{equation}
Note that
the maximum mass flux
only depends on grain properties
through $ Q\sub{pr} $ and $ \beta $.\footnote{
While larger grains constitute more mass for a given geometrical optical depth profile
($m / \tau\sub{geo} \propto m / \sigma \propto s$),
they also migrate slower ($\dot{r}\sub{PR} \propto \beta \propto s^{-1}$).
These two effects cancel each other.
}
The radiation pressure efficiency $ Q\sub{pr} $ must obey $0 \leq Q\sub{pr} \leq 2$,
and particles with $ \beta > 1 $ are always unbound.
Therefore, Eq.~\ref{eq:max_mass_flux_r0} with $ Q\sub{pr} = 2 $ and $ \beta = 1 $ gives
a solid upper limit on the inward mass flux due to \mbox{P--R} drag,
unless one of the model assumptions does not hold
(e.g., collisions are non-destructive).

\subsection{\mbox{P--R} drag and sublimation}
\label{s:prsubl}

In the preceding, we found that \mbox{P--R} drag
supplies
a small but non-zero
amount of dust to the innermost parts of a debris disk.
As this material approaches the central star, it is heated by stellar radiation.
Eventually, the dust grains become so hot that they start sublimating.
We now review the evolution of these particles considering \mbox{P--R} drag and sublimation, but ignoring collisions.

\subsubsection{Dust sublimation formalism}
\label{s:subl}

For a spherical dust grain in a gas-free environment,
the rate at which the grain radius $ s $ changes
is given by \citep[e.g.,][]{2008Icar..195..871K}
\begin{equation}
  \label{eq:dsdt}
  \frac{ \dif s }{ \dif t }
    = - \frac{ P\sub{v}(T) }{ \rho\sub{d} }
    \sqrt{ \frac{ \mu m\sub{u} }{ 2 \pi k\sub{B} T } }.
\end{equation}
Here, $P\sub{v}$ is the phase-equilibrium vapor pressure,
$\rho\sub{d}$ is the bulk density of the dust,
$\mu$ is the molecular weight of dust molecules, $m\sub{u}$ is the atomic mass unit,
$k\sub{B}$ is the Boltzmann constant, and $T$ is the temperature of the dust.
This theoretical sublimation rate is sometimes lowered to comply with experimental results,
parameterized in a sticking efficiency or accommodation coefficient.
Here, we ignore this small effect, by assuming a sticking efficiency of unity.

The temperature dependence of $P\sub{v}$ is given by \citep{2009Icar..201..395K}
\begin{equation}
  \label{eq:p_vap}
  P\sub{v}(T) = P\sub{0} \exp{ \left( - \frac{ \mu m\sub{u} H }{ k\sub{B} T }\right) },
\end{equation}
where $P\sub{0}$ is
a normalisation constant
and $H$ is the latent heat of sublimation.
By assuming $P\sub{0}$ to be constant, we neglect a small temperature dependence beyond the exponential.

Sublimation parameters are material dependent,
and can be determined by laboratory measurements.
The material we consider in this study is carbonaceous dust.
This choice is motivated by
the proximity of hot exozodiacal dust to its host star,
which suggests a very refractory material, like carbon \citep{2013ApJ...763..119M,2013A&A...555A.146L}.
Specifically, we use the sublimation parameters of graphite,
for which many laboratory measurements are available.
For the molecular weight we use $ \mu = 36.03 $,
reflecting the fact that
graphite sublimation typically releases clusters of three carbon atoms
at the temperatures and pressures relevant to this work \citep{Abrahamson1974111}.
The parameters for \mbox{C$_3$-sublimation} are
$ P\sub{0} = \mathrm{ 2.95 \times 10^{14}~dyn~cm^{-2} }$ and
$ H = \mathrm{ 2.15 \times 10^{11}~erg~g^{-1} } $ \citep{zavitsanos:2966}.
For the bulk density of the material we use $ \rho\sub{d} = \mathrm{ 1.8~g~cm^{-3} } $.

For our analytical estimates, we approximate the grain temperature $T$ by its black-body temperature
\begin{equation}
  \label{eq:temp_bb}
  T\sub{bb} = \left( \frac{ L\sub{\star} }{ 16 \pi \sigma\sub{SB} r^2 } \right)^{1/4},
\end{equation}
where $ \sigma\sub{SB} $ is the Stefan--Boltzmann constant.
In reality, the grain temperature is a function of size.
Temperatures are generally higher than $ T\sub{bb} $ for
particles that are smaller than the typical wavelength of the stellar radiation,
because these grains do not cool efficiently.
For simplicity, we ignore this effect here,
and investigate it further
in our numerical calculations,
which use realistic grain temperatures (see Sect.~\ref{s:setup_material}).

In the black-body approximation,
the sublimation rate $\dot{s}$ becomes independent of grain size.
This leads to a simple expression for the sublimation timescale $ t\sub{subl} $, defined as
the time it takes for a spherical dust grain to disappear,
which is
\begin{equation}
  \label{eq:t_subl}
  t\sub{subl}(r) = - \frac{ s }{ \dot{s}(r) }.
\end{equation}
Note that this estimate assumes the grain temperature remains constant (at $ T\sub{bb} $) throughout the sublimation process.
As a sublimating particle becomes smaller, the black-body approximation is bound to become inaccurate.
For sufficiently large particles, however,
the true sublimation time is dominated by the black-body regime,
and Eq.~\ref{eq:t_subl} provides a good approximation.

\subsubsection{The pile-up of dust in the sublimation zone}
\label{s:pileup}

As dust grains become smaller due to sublimation, their \mbox{$\beta$-ratio} changes.
As a result of this, the interplay between \mbox{P--R} drag and sublimation can lead to
a pile-up of dust in the sublimation zone.
This phenomenon was studied in detail
by \citet{2008Icar..195..871K,2009Icar..201..395K,2011EP&S...63.1067K}.
Here, we give a brief explanation of the pile-up mechanism.

When dust grains migrate inwards due to \mbox{P--R} drag,
their temperature gradually increases.
At some point, the grains are heated to the point that sublimation becomes substantial,
and their sizes start decreasing significantly.
For particles larger than the peak wavelength of the stellar spectrum $ \lambda\sub{\star} $,\footnote{
Following Wien's displacement law, $ \lambda\sub{\star} \approx 0.5~\upmu$m for the stars considered in this research.}
the radiation pressure efficiency is roughly constant at $ Q\sub{pr} \approx 1 $,
and therefore the approximation $ \beta \propto s^{-1} $ holds (see Eq.~\ref{eq:beta}).
Hence, as the particles lose mass, radiation pressure becomes more important (relative to stellar gravity),
which has the effect of increasing the semi-major axes and eccentricities of their orbits,
compensating the decrease terms due to \mbox{P--R} drag.
This effectively slows down the inward migration of the dust grains,
leading to an accumulation of dust in the sublimation zone.
Eventually, the dust grains either sublimate completely,
or their \mbox{$ \beta $-ratios} increase to the point that they become unbound
and are blown out of the system.

Accumulation of dust occurs when the decrease of semi-major axis due to \mbox{P--R} drag is compensated by
the increase of semi-major axis due to sublimation.
This happens approximately at the radial distance
where the timescale of \mbox{P--R} drag equals that of sublimation \citep{2008Icar..195..871K}.\footnote{
More precise estimates of the pile-up distance are given by \citet{2009Icar..201..395K,2011EP&S...63.1067K}.
We use this simple approximation
to facilitate the estimate of the pile-up magnitude is Sect.~\ref{s:pile_mag}.}
We denote this distance with $ r\sub{pile} $,
and determine its value by solving
$ t\sub{PR}( r\sub{pile} ) = t\sub{subl}( r\sub{pile} ) $.
Assuming that the dust in the pile-up has the black-body temperature $ T\sub{pile} = T\sub{bb}(r\sub{pile}) $,
which holds for $ s \gtrsim \lambda\sub{\star} $,
this equation can be written as
\begin{equation}
  \label{eq:temp_pile}
  \frac{ 12 \sigma\sub{SB} }{ c^2 } \frac{ Q\sub{pr} }{ P_0 } T\sub{pile}^4
  = \exp \left( - \frac{ \mu m\sub{u} H }{ k\sub{B} T\sub{pile} }\right)
  \sqrt{ \frac{ \mu m\sub{u} }{ 2 \pi k\sub{B} T\sub{pile} } },
\end{equation}
which shows that $ T\sub{pile} $
is independent of stellar parameters, the bulk density of the dust,
and grain size ($ Q\sub{pr} $ is nearly constant for $ s \gtrsim \lambda\sub{\star} $),
only depending on material properties \citep[cf.][]{2008Icar..195..871K,2009Icar..201..395K,2011EP&S...63.1067K}.
Using $ Q\sub{pr} = 1 $ and the sublimation parameters of graphite given in Sect.~\ref{s:subl},
we numerically solve Eq.~\ref{eq:temp_pile} to find $ T\sub{pile} \approx 2020$~K,
and hence
\begin{equation}
  \label{eq:r_pile_approx}
     r\sub{pile}
     \approx 0.019
     \, \biggl( \frac{ L\sub{\star} }{ \mathrm{1~L}\sub{\odot} } \biggr)^{1/2}
     \, \biggl( \frac{ T\sub{pile} }{ \mathrm{2020~K} } \biggr)^{-2}
     \; \mathrm{AU}.  
\end{equation}
The pile-up distance is independent of grain size because
larger particles take longer to sublimate,
but also migrate slower due to \mbox{P--R} drag.
This holds as long as the black-body approximation is valid, so the grain temperature in the pile-up is independent of grain size.

\subsubsection{Conditions for dust pile-up}
\label{s:pile_cond}

\citet{2011EP&S...63.1067K} list two conditions for the accumulation of dust to be substantial:
(1)~sufficiently large values of $ \beta $ need to be reached as dust grains sublimate,
and (2)~the orbital eccentricities of the dust grains need to be sufficiently low
when they enter the sublimation zone.

Considering an inward stream of dust grains with a range of sizes,
the particles contributing the most to the pile-up are those with the highest $ \beta $.
These particles have the strongest \mbox{P--R} drag drift rates,
so per unit of time more of them arrive in the sublimation zone,
where their inward drift is canceled due to sublimation.
In addition, for the typical size distribution resulting from a collisional cascade,
the total cross-section is dominated by the smallest particles.
Since dust grains with $ \beta > 0.5 $ are typically blown out as soon as they are created,
particles that are barely bound ($ \beta \approx 0.5 $) before the onset of substantial sublimation
are the most important for the pile-up.
A requirement for dust migration to slow down is that $ \beta $ increases as the grain size decreases.
For a given system, however, $ \beta $ reaches a maximum value $ \beta\sub{max} $
at $ s \sim \lambda\sub{\star} $,
because smaller particles have a lower $ Q\sub{pr} $.
Considering that relatively high values of $ \beta $ are needed for an efficient pile-up,
low luminosity stars do not have significant pile-ups.
\citet{2009Icar..201..395K,2011EP&S...63.1067K} set the limit at $ \beta\sub{max} > 0.5 $,
and derive a lower limit on the stellar luminosity for significant pile-up,
assuming that $ Q\sub{pr} \approx 1 $ holds for $ s \gtrsim \lambda\sub{\star} $.
This limit is
   \begin{equation}
     \label{eq:lum_lim}
     L\sub{\star}
     \gtrsim 0.5
     \, \biggl( \frac{ M\sub{\star} }{ \mathrm{1~M}\sub{\odot} } \biggr)
     \, \biggl( \frac{ T\sub{\star} }{ \mathrm{ 5 \times 10^3~K} } \biggr)^{-1}
     \, \biggl( \frac{ \rho\sub{d} }{ \mathrm{ 1.0~g~cm^{-3} } } \biggr)
     \; \mathrm{L}\sub{\odot},
   \end{equation}
where $ T\sub{\star} $ is the effective temperature of the central star.

The pile-up of dust in the sublimation zone is highly dependent on
the eccentricity of the dust
as it enters the sublimation zone
\citep{2008Icar..195..871K}.
For the pile-up mechanism to produce a significant enhancement,
the eccentricity of the dust particles in the sublimation zone must be very low
\citep[$e \lesssim 10^{-2}$;][]{2008Icar..195..871K,2011EP&S...63.1067K}.
Particles with higher orbital eccentricities do not spend enough time in the sublimation zone
before they are blown out (or sublimate completely)
to contribute significantly to the dust enhancement.

When dust particles are created in collisions in the parent belt,
they are put on eccentric orbits with their periastron in the parent belt.
Particles released from circular orbits will acquire orbital eccentricities of
\begin{equation}
  \label{eq:e_beta}
  e = \frac{ \beta }{ 1 - \beta }.
\end{equation}
The particles that are the most important for the pile-up
are the ones with $\beta \approx 0.5$.
These barely bound particles
initially follow
very elliptic orbits, with eccentricities close to unity.

The initial eccentricity of the $\beta \approx 0.5$ particles is far too high for any significant pile-up to occur.
However, as the dust grains migrate inward due to \mbox{P--R} drag,
their orbits are circularized.
The eccentricity evolution of dust particles experiencing \mbox{P--R} drag is coupled to their orbital size evolution
according to \citep{1950ApJ...111..134W}
\begin{equation}
  \label{eq:pr_ae}
  \frac{ a\sub{1} ( 1 - e\sub{1}^2 ) }{ a\sub{0} ( 1 - e\sub{0}^2 ) } = \left( \frac{ e\sub{1} }{ e\sub{0} } \right)^{4/5},
\end{equation}
where $(a\sub{0}, e\sub{0})$ are the initial semi-major axis and eccentricity, which evolve into $(a\sub{1}, e\sub{1})$.
This coupling can be used to place a lower limit on the distance of the source region,
if significant pile-up is to occur,
using the maximum allowed eccentricity in the sublimation zone for efficient pile-up.
Written in terms of periastron distance $ q = a ( 1- e ) $, Eq.~\ref{eq:pr_ae} becomes
  \begin{equation}
    \label{eq:pr_qe}
    \frac{ q\sub{1} ( 1 + e\sub{1} ) }{ q\sub{0} ( 1 + e\sub{0} ) } = \left( \frac{ e\sub{1} }{ e\sub{0} } \right)^{4/5},
  \end{equation}
Substituting $r\sub{0}$ and $r\sub{pile}$ for $q\sub{0}$ and $q\sub{1}$, respectively,
and using $e\sub{0} \approx 1$ and $e\sub{1} \lesssim 10^{-2}$,
yields a lower limit on the source radius
\begin{equation}
  \label{eq:r_pro_lim}
  r\sub{0} \gtrsim 20 r\sub{pile},
\end{equation}
for a significant enhancement in the dust density to be possible.
Other mechanisms
may help in the circularization of the orbits of small particles,
decreasing the limit on $r\sub{0}$.
An example is the drag force due to
small amounts of gas present in the disk \citep{2001ApJ...557..990T}.

\subsection{Pile-up in a collisionally active disk}
\label{s:pilecoll}

So far, we looked separately at the balance between \mbox{P--R} drag and collisions (without considering sublimation),
and at the balance between \mbox{P--R} drag and sublimation (without considering collisions).
We now investigate under what conditions this pairwise approach is justified,
and combine the previous findings to estimate the distribution of dust
in a debris disk in which all three processes are operational.

\subsubsection{Do collisions interfere with dust pile-up?}

Since collisions might interfere with the process of dust pile-up,
the results of Sect.~\ref{s:prsubl} are only valid if collisions do not play an important role
at distances where sublimation becomes significant.
We now investigate whether
collisions can indeed be neglected in the inner regions of debris disks
by comparing the characteristic timescales of the three processes as a function of radius.

\begin{figure}[!t]
  \includegraphics[width=\linewidth]{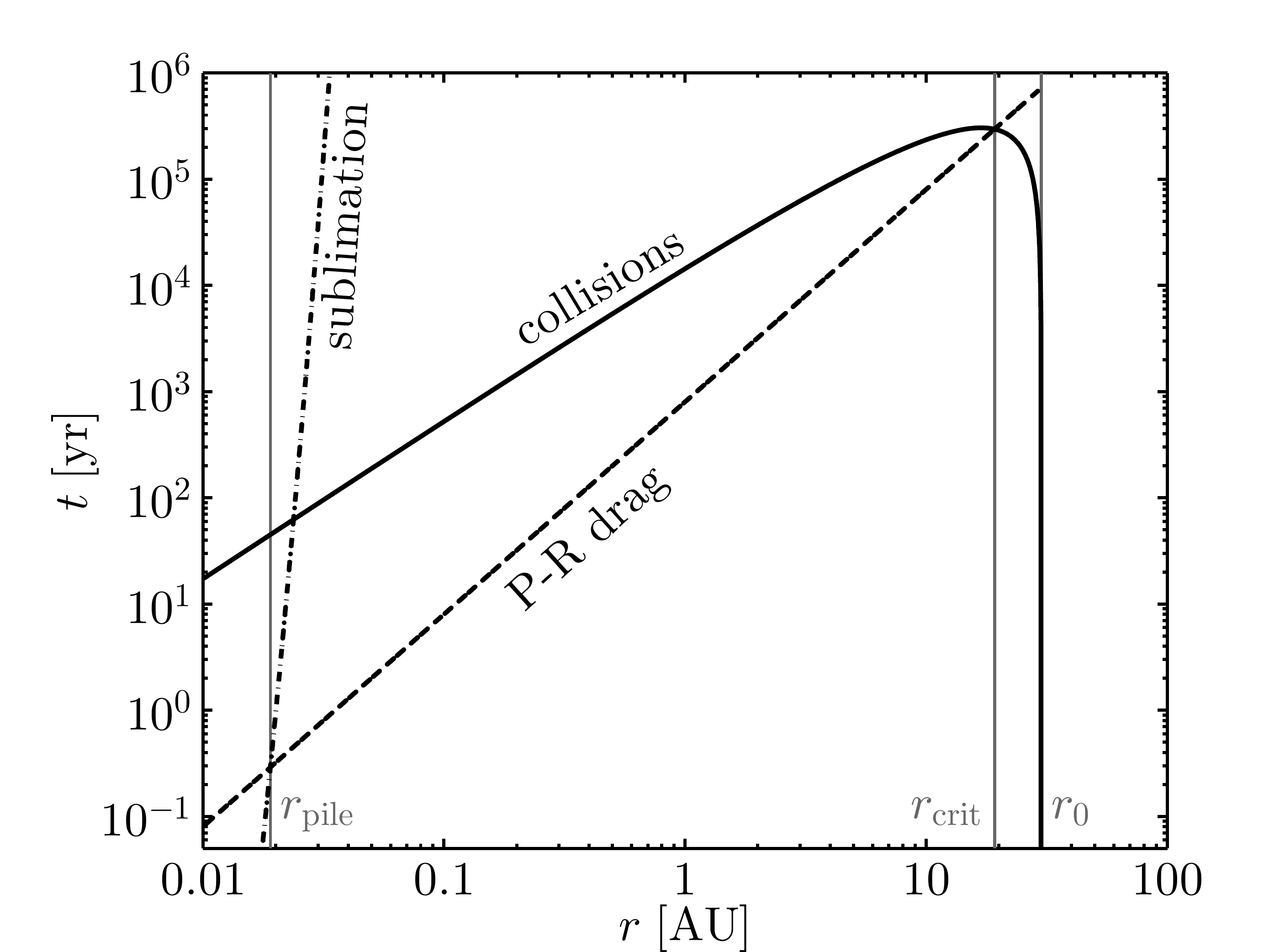}
  \caption{The characteristic timescale as function of radial distance
  for sublimation (Eq.~\ref{eq:t_subl}), \mbox{P--R} drag (Eq.~\ref{eq:t_P--R}),
  and
  mutual collisions (minimum timescale, Eq.~\ref{eq:min_t_coll}),
  for $\beta = 0.5$ particles
  in a debris disk around a solar-mass star with a very dense
  ($\eta\sub{0} \gg 1$)
  parent belt located at 30~AU.
  The grey vertical lines indicate
  the radial distances used by the analytical model:
  $ r\sub{pile} $ is the radius for which the sublimation timescale equals the \mbox{P--R} drag timescale,
  $ r\sub{ crit } $ is the radius for which the \mbox{P--R} drag timescale equals the collisional timescale,
  and $ r\sub{ 0 } $ is the location of the parent belt.
  For the sublimation timescale, we assume
  that the dust grains are
  solid spheres of graphite (see Sec.~\ref{s:subl} for the values of the sublimation parameters)
  with a radius of  $ s = \mathrm{0.64~\upmu m} $
  (the size corresponding to $\beta = 0.5$).
  }
  \label{fig:timescales}
\end{figure}

In the case of a very dense parent belt ($\eta\sub{0} \gg 1 $), $\tau\sub{geo}(r)$ is given by Eq.~\ref{eq:max_tau_eff}.
Putting this in Eq.~\ref{eq:t_coll}
results in the minimum collisional timescale
\begin{equation}
  \label{eq:min_t_coll}
  \min \left[ t\sub{coll}(r) \right] = \frac{ c r^{3/2} }{ \beta G M\sub{\star} } \left( \sqrt{ r\sub{0} } - \sqrt{ r } \right),
  \qquad r \leq  r\sub{0}.
\end{equation}
 In Fig.~\ref{fig:timescales}
this timescale is compared with the sublimation and \mbox{P--R} drag timescales
for a debris disk around a solar-mass star with a dense parent belt at 30~AU,
consisting of barely bound ($ \beta = 0.5 $) dust grains,
using the sublimation parameters of graphite.
In this example system,
the collisional timescale is longer than the other timescales at $ r\sub{pile} $,
implying that
\mbox{P--R} drag dominates over collisions in the sublimation zone,
and collisions do not interfere with dust pile-up.
We now check if this a general result, or under which conditions it is the case.

Let $r\sub{ crit }$ be the radial distance at which collisional and \mbox{P--R} drag timescales are equal.
Solving $ t\sub{PR}( r\sub{ crit } ) = t\sub{coll}( r\sub{ crit } ) $ for $ r\sub{ crit } $,
with $\tau\sub{geo}(r)$ given by Eq.~\ref{eq:tau_eff_prcoll}, yields
\begin{equation}
  \label{eq:r_crit}
  \frac{ r\sub{ crit } }{ r\sub{0} } = \left( \frac{ 4 \eta\sub{0} + 1 }{ 5 { \eta\sub{0} } } \right)^2.
\end{equation}
For $ \eta\sub{0} = 1 $ we recover $r\sub{ crit } = r\sub{0}$, as expected.
From the limiting case $\eta\sub{0} \gg 1 $ (i.e., a very dense parent belt), we find
\begin{equation}
  \label{eq:r_crit_lim}
  r\sub{ crit } > 0.64 r\sub{0}.
\end{equation}
This shows that far enough inward from the parent belt,
\mbox{P--R} drag always dominates over collisions.
Furthermore, if Eq.~\ref{eq:r_pro_lim} holds
we find $r\sub{ crit } \gtrsim 12.8 r\sub{pile}$,
which means that in systems where the parent belt is distant enough for significant dust pile-up to occur,
\mbox{P--R} drag dominates over collisions in the sublimation zone,
and collisions are so infrequent there that they do not interfere with the pile-up process.

The pile-up of dust means that $\tau\sub{geo}$ increases around $ r\sub{pile} $,
locally decreasing the collisional timescale.
However, a significant pile-up requires Eq.~\ref{eq:r_pro_lim} to hold,
in which case the ratio between the minimum collisional timescale
and the \mbox{P--R} drag timescale is found to satisfy
$ \min \left[ t\sub{coll}(r\sub{pile}) \right] / t\sub{PR}(r\sub{pile}) =
4 ( \sqrt{ r\sub{0} / r\sub{pile} } - 1 ) \gtrsim 13.9 $.
To overcome this difference, the pile-up would have to raise $\tau\sub{geo}$ by the same factor.
Since the \mbox{$\tau\sub{geo}$-enhancement} factor is never found to be greater than about 10 \citep{2009Icar..201..395K},
we expect the disk to remain drag (or sublimation) dominated inside $r\sub{ crit }$.

\subsubsection{Estimating the pile-up magnitude}\label{s:pile_mag}

The efficiency of dust pile-up was studied in detail by \citet{2009Icar..201..395K,2011EP&S...63.1067K},
who give formulae for the resulting enhancement in particle number density and geometrical optical depth.
Here, we present
a simple order of magnitude estimate of the maximum amount of material in the pile-up,
which can be used to assess whether pile-ups can explain the observed NIR excess emission.

Dust grains reside in the pile-up for roughly one sublimation timescale,
after which they are either completely sublimated,
or their size is reduced so much that they are blown out of the system.
Since the pile-up occurs roughly at $ r\sub{pile} $,
defined such that $ t\sub{subl}( r\sub{pile} ) = t\sub{PR}( r\sub{pile} ) $,
the dust stays in the pile-up for about one \mbox{P--R} drag timescale.
Given this, the total number of particles in the pile-up is
\begin{equation}
 \label{eq:n_pile}
 N\sub{pile}
 = \varphi\sub{PR}( r\sub{pile} ) t\sub{PR}( r\sub{pile} ).
\end{equation}

To describe the radial profile in terms of geometrical optical depth,
it is necessary to specify the radial width of the pile-up $ \Delta r\sub{pile} $.
In reality, $ \Delta r\sub{pile} $ depends on the orbital eccentricities of the particles in the pile-up,
and on differences in pile-up distance for particles of different sizes that contribute.
Since this is beyond the scope of this work,
we keep the relative pile-up width $ \Delta r\sub{pile} / r\sub{pile} $ as a free parameter.

\begin{figure}[!t]
  \includegraphics[width=\linewidth]{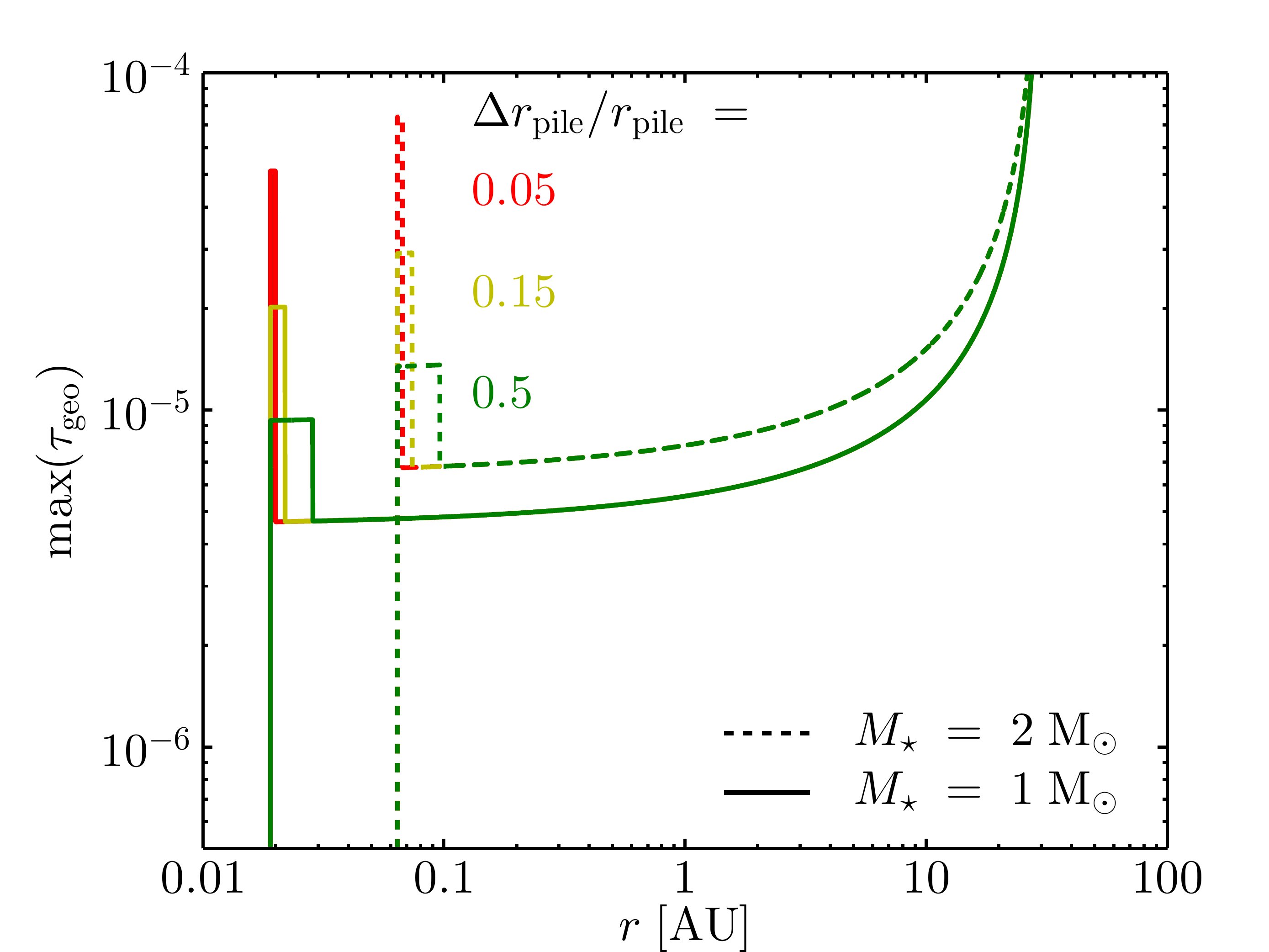}
  \caption{
  Maximum geometrical optical depth profiles
  of debris disks with collisions, \mbox{P--R} drag, and sublimation
  (Eq.~\ref{eq:tau_full} with $ \tau\sub{geo,\,base}(r) $ given by Eq.~\ref{eq:max_tau_eff}),
  for different values of relative pile-up width $ \Delta r\sub{pile} / r\sub{pile} $ shown in different colors.
  The solid lines correspond to disks around solar-mass stars, the
  dashed lines to disks around 2~M$\sub{\odot}$ stars.
  All profiles assume a parent belt at $ r\sub{0} = 30$~AU, dust grains with $ \beta = 0.5 $,
  and $ r\sub{pile} \approx 0.019$~AU, and $ r\sub{pile} \approx 0.064$~AU,
  for the $ M\sub{\star} = \mathrm{1~M}\sub{\odot} $,
  and $ M\sub{\star} = \mathrm{2~M}\sub{\odot} $ case, respectively,
  corresponding to spherical graphite grains with $ \beta = 0.5 $.
  }
  \label{fig:max_tau_pileup}
\end{figure}

Combining Eqs.~\ref{eq:tau_eff_approx}, \ref{eq:ndot_pr}, and \ref{eq:n_pile},
we find that the geometrical optical depth due to the material in the pile-up is given by
\begin{equation}
  \label{eq:tau_pile}
  \tau\sub{geo,\,pile}
  = \frac{ \sigma N\sub{pile} }{ 2 \pi r\sub{pile} \Delta r\sub{pile} }
  = \frac{ r\sub{pile} }{ 2 \Delta r\sub{pile} } \tau\sub{geo,\,base}( r\sub{pile} ),
\end{equation}
where $ \tau\sub{geo,\,base}( r ) $ denotes
the base level of dust
due to \mbox{P--R} drag and collisions
given by Eq.~\ref{eq:tau_eff_prcoll}.
\citet{2009Icar..201..395K} find that sublimating dust particles slowly move outward.
Therefore, we assume that the pile-up extends from $ r\sub{pile} $ outwards,
overlapping with the inward migrating material.
The complete
geometrical optical depth profile,
which includes the effects of collisions, \mbox{P--R} drag, and sublimation,
can then be formally described by \citep[cf.][]{2011EP&S...63.1067K}
\begin{equation}
\label{eq:tau_full}
\tau\sub{geo}(r) =
  \begin{cases}
     0 & \text{for } r <  r\sub{pile} \\
     \tau\sub{geo,\,base}(r) + \tau\sub{geo,\,pile} & \text{for } r\sub{pile} \leq r \leq r\sub{pile} + \Delta r\sub{pile} \\
     \tau\sub{geo,\,base}(r) & \text{for } r\sub{pile} + \Delta r\sub{pile} < r \leq r\sub{0}
  \end{cases}\!.
\end{equation}
Using Eq.~\ref{eq:max_tau_eff} for $ \tau\sub{geo,\,base}(r) $ gives the maximum profile.
Figure~\ref{fig:max_tau_pileup} shows this maximum profile for different values of $ \Delta r\sub{pile} / r\sub{pile} $.

We can make a rough estimate of the     extbf{geometrical} optical depth enhancement factor $ f\sub{\tau_{geo}} $ of the pile-up
(i.e., how much higher $ \tau\sub{geo} $ in the pile-up is compared to the base level of the inner disk),
as function
of the radial width of the pile-up $ \Delta r\sub{pile} $.
\begin{equation}
  \label{eq:enhance_tau}
  f\sub{\tau\sub{geo}}
  = \frac{ \tau\sub{geo,\,pile} + \tau\sub{geo,\,base} }{ \tau\sub{geo,\,base} }
  = \frac{ r\sub{pile} }{ 2 \Delta r\sub{pile} } + 1.
\end{equation}
For a pile-up width of $ \Delta r\sub{pile} \approx 0.05 r\sub{pile} $ \citep{2011EP&S...63.1067K},
this gives $ f\sub{\tau_{geo}} \approx 11 $,
which is comparable to the highest values found by \citet[][see their Fig.~6]{2009Icar..201..395K}
for carbonaceous dust grains around early \mbox{F-type} stars.

For comparison with observations, it suffices to assume that
all the pile-up material is located at $r\sub{pile}$ (i.e., $ \Delta r\sub{pile} / r\sub{pile} \ll 1 $).
This gives the highest possible temperature to all pile-up particles,
and therefore results in the maximum NIR flux.
In the \mbox{$ \tau\sub{geo} $-profile}, however, it would give a singularity at $ r = r\sub{pile} $.
To avoid this, we instead compute the fractional luminosity of the pile-up.
Fractional luminosity is defined as
the ratio of the infrared luminosity of the dust $ L\sub{D} $ to the stellar luminosity.
Assuming the disk is radially optically thin
and the dust grains have unity absorption and emission efficiencies at all wavelengths (i.e., assuming black-body grains),
it can be approximated
by the fraction of the star that is covered by dust
\begin{equation}
  \label{eq:fraclum}
  \frac{ L\sub{D} }{ L\sub{\star} } = \int \frac{ \sigma n(r) }{ 4 \pi r^2 } \, \dif r .
\end{equation}
Evaluating this with $ n(r) \dif r = N\sub{pile} $,
and using the maximum geometrical optical depth (Eq.~\ref{eq:max_tau_eff}),
we find that the maximum fractional luminosity due to material in the pile-up is
\begin{equation}
  \label{eq:max_fraclum_pr}
  \max \left[ \left( \frac{ L\sub{D} }{ L\sub{\star} } \right)\sub{pile} \right]
  = \frac{ \sqrt{ G M\sub{\star} } \, \beta }{ 8 c \left( \sqrt{ r\sub{0} } - \sqrt{ r\sub{pile} } \right) }.
\end{equation}
In the limit of $ r\sub{0} \gg r\sub{pile} $, this becomes
\begin{equation}
  \label{eq:max_fraclum_pr_r0}
  \max \left[ \left( \frac{ L\sub{D} }{ L\sub{\star} } \right)\sub{pile} \right]
  \approx 6.2 \times 10^{-6}
  \, \biggl( \frac{ M\sub{\star} }{ \mathrm{1~M}\sub{\odot} } \biggr)^{1/2}
  \, \biggl( \frac{ r\sub{0} }{ \mathrm{1~AU} } \biggr)^{-1/2}
  \, \biggl( \frac{ \beta }{ 0.5 } \biggr).
\end{equation}
Note that this is only the fractional luminosity due to the dust in the pile-up.
Material just beyond
$r\sub{pile}$
is not accounted for,
and will increase the total fractional luminosity.

\subsection{Summary of analytical findings}
\label{s:analitic_sum}

The analytical model presented above
yields several tentative conclusions about dust in the inner parts of debris disks:
\begin{enumerate}
  \item \mbox{P--R} drag gives rise to a small but non-zero inward mass flux of dust in the inner disk,
        which is self-limited by collisions (Eq.~\ref{eq:max_mass_flux_r0}).
  \item A pile-up of sublimating dust occurs,
        as long as the star is luminous enough (Eq.~\ref{eq:lum_lim}),
        and the parent belt is distant enough (Eq.~\ref{eq:r_pro_lim}).
  \item \mbox{P--R} drag dominates over collisions in the inner parts of the disk (Eq.~\ref{eq:r_crit_lim}),
        so collisions do not interfere with dust pile-up.
  \item Given that the pile-up of dust occurs around the radial distance
        where the sublimation timescale equals the \mbox{P--R} drag timescale,
        and that sublimating dust resides in the pile-up for about one sublimation timescale,
        there is a maximum fractional luminosity that this dust can provide (Eq.~\ref{eq:max_fraclum_pr_r0}).
\end{enumerate}

\section{Numerical modeling}
\label{s:numeric}

The analytical approach used in Sect.~\ref{s:analytic} contains several simplifying assumptions.
Most importantly, we only self-consistently solve the continuity equation for \mbox{P--R} drag and collisions
(under the assumptions listed in Sect.~\ref{s:prcoll}),
and afterwards estimate the effect of dust pile-up due to sublimation,
assuming the grains reside in the sublimation zone for one \mbox{P--R} drag timescale (see Sect.~\ref{s:pile_mag}).
To test the impact of these assumptions, we now proceed to solve the continuity equation numerically
using a debris disk model that self-consistently handles the effects of
stellar gravity, direct radiation pressure, \mbox{P--R} drag, sublimation, and destructive collisions.
Our strategy here is to simulate a few specific cases,
and compare the results to the analytical maximum distributions found in Sect.~\ref{s:analytic},
to assess the validity and generality of these simple expressions.
A description of our numerical debris disk model is given in Sect.~\ref{s:model},
the runs we performed are detailed in Sect.~\ref{s:setup},
and the resulting dust distributions are presented in Sect.~\ref{s:results}.

\subsection{Model description}
\label{s:model}

Our debris disk model closely follows the method developed by \citet{2005Icar..174..105K,2006A&A...455..509K} and \citet{2008PhDTLohne}.
We refer the reader to these publications for a detailed description of the method,
and only provide a brief outline of its principles here.
We focus on the 
changes we made,
which are the inclusion of a time-dependent treatment of dust sublimation (see Sect.~\ref{s:model_subl}),
and the implementation of additional numerical acceleration techniques (see Appendix~\ref{s:app_num}).
Our code was tested by
comparing its predictions to solutions of the equations of motion and sublimation for individual particles,
and by benchmarking it against
the results of \citet{2006A&A...455..509K}.
This verification is described in Appendix~\ref{s:app_verif}.
Two physical processes that are included by \citet{2008PhDTLohne},
but not
considered by
our present code, are
stellar wind drag and erosive (cratering) collisions.
We discuss the impact
they
may have on our results in Sect.~\ref{s:discussion}.

\subsubsection{Method basics}
\label{s:model_basics}

The method of \citet{2005Icar..174..105K} applies the kinetic method of statistical physics to debris disks,
simultaneously following
the spatial and size distributions of
dust and planetesimals
in a phase space of orbital elements and particle masses.
Using orbital element instead of radial distance to follow the spatial distribution
makes it possible to account for particles on eccentric orbits,
whose orbits can span a large range of radial distances.
The continuity equation is solved in this phase space,
with processes that affect the evolution of a particle's phase-space coordinates
in a continuous fashion (\mbox{P--R} drag and sublimation) as diffusion terms,
and processes that abruptly change phase-space coordinates (collisions)
as source and sink terms.
Formally, this is decribed by \citep[cf.][]{2005Icar..174..105K,2008PhDTLohne}
\begin{equation}
  \label{eq:master}
  \frac{ \dif n }{ \dif t } ( m, \vec{k}, t )
    = \left( \frac{ \dif n }{ \dif t } \right)\sub{source}
    - \left( \frac{ \dif n }{ \dif t } \right)\sub{sink}
    - \mathrm{div} \left( n \frac{ \dif \{ m, \vec{k} \} }{ \dif t } \right),
\end{equation}
where $ n ( m, \vec{k}, t ) $ is the phase-space number density
at time $ t $
(i.e., the distribution function that describes the state of the disk),
$ \vec{k} $ is the vector of orbital elements,
and $\{ m, \vec{k} \}$ denotes the vector consisting of $m$ and $\vec{k}$.
The divergence term represents the diffusion of material in phase space
(i.e., transport due to \mbox{P--R} drag and sublimation).
For brevity, we omit the
arguments $ ( m, \vec{k}, t ) $ for
all terms on the right hand side of the equation.

To make the
numerical evaluation of the continuity equation
manageable with limited computational capacity,
the number of phase-space variables needs to be reduced.
By assuming the disk is axisymmetric, the distribution function can be averaged
over three of the orbital elements: longitude of the ascending node, argument of the periastron, and true anomaly.
This implicitly assumes that collisional timescales are much longer than orbital timescales,
which generally holds for debris disks \citep[for a more detailed discussion, see Sect.~3.1.3 of][]{2008PhDTLohne}.
A further assumption is
that the distribution of particles over inclinations is constant,
which allows the averaging of the distribution function over inclination.
The three remaining phase-space variables
are (1)~particle mass $m$, (2)~orbital eccentricity $e$,
and (3)~an orbital element characterizing the size of the orbit,
such as semi-major axis $a$, or periastron distance $q = a ( 1 - e )$.

The final phase-space variable
can be chosen to fit the numerical needs of the problem under investigation.
For our study of the pile-up of dust due to sublimation, we choose periastron distance.
A particle on an eccentric orbit experiences most sublimation around the periastron,
due to the strong temperature dependence of sublimation.
Since the periastron distance does not evolve for \mbox{$\beta$-changes} that happen at the periastron,
the orbit's periastron distance changes much slower than its semi-major axis.
Hence, using $q$ instead of $a$ as phase-space variable
has numerical advantages.

In practice, the phase space is divided into a grid of bins,
and the distribution function
is replaced by a vector
listing the number of particles in each bin.
Equation~\ref{eq:master} then becomes a system of ordinary differential equations.
The source, sink, and diffusion terms are discretized,
and determine the rates at which the particle numbers evolve,
dependent on the population levels of other bins.
We now proceed to describe these terms
for each of the physical processes considered by our model.

\subsubsection{Poynting--Robertson drag}
\label{s:model_prdrag}

\mbox{P--R} drag affects the orbits of particles, circularizing them, and making them smaller.
These effects are accounted for in the model by diffusion terms in the continuity equation
that move particles to adjacent bins in the phase-space grid.
Since the \mbox{P--R} drag timescale is usually longer than the orbital period,
we use
the orbit-averaged change rates of the orbital elements, given by \citep[e.g.,][]{1979Icar...40....1B}
\begin{align}
  \label{eq:pr_a}
    \left< \frac{ \dif a }{ \dif t } \right> \sub{PR} & =
      - \frac{ \beta G M\sub{\star} }{ c a }
      \frac{ 2 + 3 e^2 }{ ( 1 - e^2 )^{3/2} }, \\
  \label{eq:pr_e}
    \left< \frac{ \dif e }{ \dif t } \right> \sub{PR} & =
      - \frac{ 5 \beta G M\sub{\star} }{ 2 c a^2 }
      \frac{ e }{ ( 1 - e^2 )^{1/2} }.
\end{align}
For the rate of change in periastron distance, we find
\begin{align}
  \left< \frac{ \dif q }{ \dif t } \right> \sub{PR}
    & = \frac{ \partial q }{ \partial a } \left< \frac{ \dif a }{ \dif t} \right> \sub{PR}
      + \frac{ \partial q }{ \partial e } \left< \frac{ \dif e }{ \dif t} \right> \sub{PR} \\
  \label{eq:pr_q}
    & = - \frac{ \beta G M\sub{\star} }{ 2 c q }
    \frac{ ( 4 - e ) ( 1 - e )^3 }{ ( 1 - e^2 )^{3/2} }.
\end{align}

\subsubsection{Sublimation}
\label{s:model_subl}

The formalism that we use for dust sublimation is described in Sect.~\ref{s:subl}.
For a spherical dust particle, it gives a mass loss rate of
\begin{equation}
  \label{eq:dmdt}
  \frac{ \dif m }{ \dif t }
    = - P\sub{v}(T) s^{2}
    \sqrt{ \frac{ 8 \pi \mu m\sub{u} }{ k\sub{B} T } }.
\end{equation}
In our numerical model, we use realistic dust grain temperatures
(as opposed to the black-body temperatures used in Sect.~\ref{s:analytic}).
The method for computing these temperatures is explained in Sect.~\ref{s:setup_material}.

Since the sublimation rate is strongly dependent on grain temperature,
which varies along the path of an eccentric orbit,
the mass loss rate needs to be averaged over the orbit.
As described qualitatively in Section~\ref{s:pileup},
the change in \mbox{$ \beta $-ratio} associated with mass loss
induces changes in orbital elements.
\citet{2009Icar..201..395K} derive the orbit-averaged change rate in orbital elements and mass to be
\begin{align}
  \label{eq:dadt_subl}
  \left< \frac{ \dif a }{ \dif t } \right> \sub{subl}
    & = - \frac{ \dif \ln \beta }{ \dif \ln m }
      \left( \frac{ 1 + e^2 }{ 1 - e^2 } \bar{\psi}\sub{m}
      + \frac{ 2e }{ 1 - e^2 } \bar{\phi}\sub{m} \right)
      \frac{ \beta }{ 1 - \beta } \frac{ a }{ m }, \\
  \label{eq:dedt_subl}
  \left< \frac{ \dif e }{ \dif t } \right> \sub{subl}
    & = - \frac{ \dif \ln \beta }{ \dif \ln m }
      ( e \bar{\psi}\sub{m} + \bar{\phi}\sub{m} )
      \frac{ \beta }{ 1 - \beta } \frac{ 1 }{ m }, \\
  \label{eq:dmdt_mean}
  \left< \frac{ \dif m }{ \dif t } \right> \sub{subl}
    & = - \bar{\psi}\sub{m},
\end{align}
with
\begin{align}
  \label{eq:psi_bar_m}
  \bar{\psi}\sub{m}
  & = - \frac{ 1 }{ 2 \pi } \int \limits_{ 0 }^{ 2 \pi } \frac{ \dif m }{ \dif t }
  \frac{ ( 1 - e^2 )^{3/2} }{ (1 + e \cos f )^2 } \, \dif f, \\
  \label{eq:phi_bar_m}
  \bar{\phi}\sub{m}
  & = - \frac{ 1 }{ 2 \pi } \int \limits_{ 0 }^{ 2 \pi } \frac{ \dif m }{ \dif t }
  \cos f \frac{ ( 1 - e^2 )^{3/2} }{ (1 + e \cos f )^2 } \, \dif f,
\end{align}
where $f$ denotes the true anomaly.
For the periastron distance, we find
\begin{align}
  \left< \frac{ \dif q }{ \dif t } \right> \sub{subl}
    & = \frac{ \partial q }{ \partial a } \left< \frac{ \dif a }{ \dif t} \right> \sub{subl}
      + \frac{ \partial q }{ \partial e } \left< \frac{ \dif e }{ \dif t} \right> \sub{subl} \\
  \label{eq:dqdt_subl}
    & = - \frac{ \dif \ln \beta }{ \dif \ln m }
      \frac{ \bar{\psi}\sub{m} - \bar{\phi}\sub{m} }{ 1 + e }
      \frac{ \beta }{ 1 - \beta } \frac{ q }{ m },
\end{align}
For each phase-space bin, quantities $ \bar{\psi}\sub{m} $ and $ \bar{\phi}\sub{m} $
are numerically evaluated using the standard Euler method.\footnote{
Integrating over true anomaly rather than, e.g., mean anomaly warrants
a higher sampling around the periastron, where sublimation rate varies the most.}
The change rates of the phase-space variables
(Eqs.~\ref{eq:dedt_subl}, \ref{eq:dmdt_mean}, and \ref{eq:dqdt_subl})
are then used in diffusion terms in the continuity equation.

Using orbit averaged mass loss rates is only correct if
the sublimation timescale is longer than the orbital period.
For phase-space bins for which this does not hold (small particles close to the star),
we compute an equilibrium population of particles from
the product of their sublimation timescale
and the sum of their gain terms.
This implicitly assumes that particles are created on their orbits
with a uniform distribution over true anomaly.

\subsubsection{Collisions}
\label{s:model_coll}

Collisions are different from \mbox{P--R} drag and sublimation
in that they cause abrupt, rather than smooth, changes in phase-space coordinates.
In the continuity equation,
they are described by sink terms at the phase-space coordinates of targets and projectiles,
and by source terms at the coordinates of the resulting fragments.
Here, we give a summary of the way our model handles collisions.
More thorough descriptions of the treatment of collisions,
including all relevant equations, are given by \citet{2006A&A...455..509K} and \citet{2008PhDTLohne}.

For each pair of phase-space bins, we determine collision rates
for a range of relative orbit orientations (differences in the longitudes of the periastra of the two orbits).
The collision rate is
the product of
the target and projectile
number densities,
their relative velocity,
the collisional cross-section,
and the effective volume of interaction.
\citet{2006A&A...455..509K} found analytical expressions for these factors in two dimensions
as function of the orbital elements and masses corresponding to both bins, and
the relative orientation of the orbits.
To account for the third dimension, a correction is applied based on the semi-opening angle of the disk,
equivalent to the maximum inclinations of the particles.
This correction assumes that the disk is relatively flat,
consistent with observations of resolved edge-on systems.

To save computational power, we ignore collisions involving unbound particles.
This is a valid approximation
if the radial geometrical optical depth of the system is much smaller than unity
(true for most debris disks),
because then
the blowout timescale of such particles is much shorter than their collisional timescale.\footnote{
If the radial geometrical optical depth is higher than $\sim$10$^{-2}$,
the disk is subject to dust avalanches \citep{2007A&A...461..537G}.}

The nature of a collision is determined by the impact energy available per unit of mass.
We only consider catastrophic collisions,
defined as destructive events in which
the largest fragment contains at most half of the mass of the more massive of the two impactors.
The threshold for these catastrophic collisions is the critical specific energy for dispersal $ Q^\star\sub{D} $, which
incorporates the fact that fragments may reassemble after destruction, and
generally depends on particle size.
If the specific energy of a collision is higher than this threshold,
the impact destroys both bodies, and their mass is distributed over a swarm of fragments.
At specific energies just below $ Q^\star\sub{D} $, collisions are erosive.
Such cratering collisions, however, are not considered in our present model.
Hence,
if the specific energy of an impact is lower than
$ Q^\star\sub{D} $, no collision is considered to occur.

A catastrophic collision results in a range of fragments with different masses and orbits.
The fragments are distributed over particle masses according to
a single power law,
up to a maximum fragment mass, which is determined by
the kinetic energy of the impact and the material strength of the target.
The  maximum fragment mass is at most half of the mass of the target and projectile combined,
but it can also be less, if the specific energy involved in the collision is more than $ Q^\star\sub{D} $.
The amount of particles that end up in each mass bin (up to the maximum fragment mass)
is computed by integrating the fragment mass distribution.
Particles with masses below the lowest mass bin (i.e., that fall off the grid)
are considered lost due to immediate blowout or vaporization.
For each fragment mass bin, new orbital elements are calculated
using the conservation of momentum,
and taking into account direct radiation pressure (i.e., the values of $\beta$ of the fragments),
using Eqs. 19 and 20 of \citet{2006A&A...455..509K}.
This assumes that the fragments are not launched away from the collision with any velocity.
These orbital elements are rounded to find the periastron distance and eccentricity bin
corresponding to each fragment mass bin.

\subsection{Setup of the model runs}
\label{s:setup}

\subsubsection{Stellar and disk parameters}

Hot exozodiacal dust has been detected around stars with spectral types ranging from A to K (see Tbl.~\ref{tbl:obs}).
To focus on this range of stellar types,
we did one model run with a solar-mass star
and one using a $ \mathrm{2~M}\sub{\odot} $ star.
Following the mass--luminosity relation for main-sequence stars
\citep[$ L\sub{\star} \propto {M\sub{\star}}^{3.5} $;][]{1976asqu.book.....A},
we set the stellar luminosities corresponding to these stellar masses to
$ L\sub{\star} = \mathrm{1~L}\sub{\odot} $ and $ L\sub{\star} = \mathrm{11.31~L}\sub{\odot} $, respectively.

Both runs use a parent belt radius of $ r\sub{0} = 30 $~AU.
Lower values of $ r\sub{0} $ can in principle yield higher dust levels in the innermost regions (see Eq.~\ref{eq:max_tau_eff}),
but parent belts closer to the star are generally not dense enough to provide these large amounts of dust,
because they do not survive the intense collisional grinding \citep{2003ApJ...598..626D,2007ApJ...658..569W}.
In addition, many of the observed outer belts are located at tens of AUs (see Tbl.~\ref{tbl:obs}).
The level of dust in the source region
is set to $ \tau\sub{geo} ( r\sub{0} ) \approx 5 \times 10^{-5} $,
chosen such that (i.e., iterated until) the geometrical optical depth in the inner regions
does not become any higher
by increasing the level of dust at the source.\footnote{
The actual input parameter used
(which indirectly determines the geometrical optical depth at the source region)
is the mass supply term of large dust particles in the source region (see Sect.~\ref{s:sim_strategy}).
It is set to $ 10^{-10} \; \mathrm{M}\sub{\earth} \; \mathrm{yr}^{-1} $
for the $ M\sub{\star} = \mathrm{1~M}\sub{\odot} $ run,
and $ 8 \times 10^{-10} \; \mathrm{M}\sub{\earth} \; \mathrm{yr}^{-1} $
for the $ M\sub{\star} = \mathrm{2~M}\sub{\odot} $ run.
Comparing these mass fluxes to those given by Eq.~\ref{eq:max_mass_flux_r0}
indicates that the vast majority of the material is destroyed in collisions before it reaches the sublimation zone.
}
This roughly corresponds to $ \eta\sub{0} \sim 10 $ for both stellar mass cases,
which is apparently
sufficient to approximate
an inner disk \mbox{$ \tau\sub{geo} $-profile} corresponding to $ \eta\sub{0} \gg 1 $.
For the semi-opening angle of the disk we use $ \varepsilon = 8.5\degr$.

\subsubsection{Material properties}
\label{s:setup_material}

We consider carbonaceous dust particles with a density of $ \rho\sub{d} = \mathrm{ 1.8~g~cm^{-3} } $.
To compute the optical properties of these grains with different radii, we use
the DHS method of \citet{2005A&A...432..909M} with an irregularity parameter of $f_\mathrm{max}=0.8$.
This method simulates the properties of irregularly shaped particles.
For the material we use amorphous carbon with the refractive index data taken from \citet{1993A&A...279..577P}.
These optical properties are used to compute $\beta(s)$ and dust temperatures.
The dust temperatures are computed by solving the balance between absorption and thermal emission
as a function of grain size and distance to the central star.
For the sublimation properties, we use those of graphite, given in Sect.~\ref{s:subl}.

The modeling of collisions requires a prescription for the specific energy threshold for dispersal $ Q\sub{D}^{\star} $,
which is generally found to be size-dependent.
In our numerical simulation we consider particles with radii up to 1~cm. (see Sect.~\ref{s:sim_strategy}).
For bodies smaller than $s \sim 100$~m,
$ Q\sub{D}^{\star} $ is often described by
a power law with a negative exponent \citep[e.g.,][]{1999Icar..142....5B}.
However, such a prescription predicts unrealistically high values for the small particles that we consider.
Therefore, following \citet{2010MNRAS.401..867H},
we use the constant value of $ Q\sub{D}^{\star} = 10^{-7}$~erg~g$^{-1}$ found in
laboratory experiments with high-velocity collisions of small particles \citep{2004P&SS...52.1129F}.
We follow \citet{2006A&A...455..509K} in setting
the collisional fragment mass distribution to $ n(m) \propto m^{-11/6} $,
and assuming the maximum fragment mass scales with specific impact energy to the power $-1.24$ \citep[][]{1977Icar...31..277F}.

\subsubsection{The phase-space grid}

Because this problem is computationally very demanding,
great care has to be taken in
setting up the phase-space grid.
Specifically, resolving the pile-up requires a high resolution
in the sublimation zone and at small particles sizes, where radiation pressure becomes important.
To achieve this with limited computational resources,
we designed a non-uniform grid that has a higher resolution where it is required.

The eccentricity grid contains 10 logarithmic bins between $ e = 0 $ and $ e = 1 $,
with the lowest bin at $ e = 10^{-4} $.
In addition, there are 2 linearly spaced eccentricity bins between $ e = 1 $ and $ e = 2 $ for hyperbolic orbits,
as well as 2 bins between $ e = -2 $ and $ e = -1 $
to account for ``anomalous'' hyperbolic orbits followed by $ \beta > 1 $ particles \cite[see][]{2006A&A...455..509K}.

The periastron distance grid consists of
two parts:
(1)~A high-resolution, linear grid of 21 bins is used to cover the sublimation zone
($ \mathrm{0.01~AU} < q < 0.03 $~AU for the $ M\sub{\star} = \mathrm{1~M}\sub{\odot} $ run,
and $ \mathrm{0.05~AU} < q < 0.1 $~AU for the $ M\sub{\star} = \mathrm{2~M}\sub{\odot} $ run).
(2)~A low-resolution, logarithmic grid covers the rest of the disk, out to about $ q = 100 $~AU,
with 60 bins in the $ M\sub{\star} = \mathrm{1~M}\sub{\odot} $ run,
and 50 bins in the $ M\sub{\star} = \mathrm{2~M}\sub{\odot} $ run.
Care was taken to place one bin exactly at $ q = 30$~AU,
which was used as the source region.

The mass grid has 48 logarithmically spaced bins in both runs,
with a higher resolution at the smaller sizes ($ \beta \gtrsim 0.05 $),
and a maximum mass corresponding to $s = 1$~cm.
For the $ M\sub{\star} = \mathrm{1~M}\sub{\odot} $ run,
the high-resolution part consists of 30 bins between $s = 0.5~\upmu$m and $s = 10~\upmu$m.
The high-resolution part of the $ M\sub{\star} = \mathrm{2~M}\sub{\odot} $ run has 36 bins
between $s = 2~\upmu$m and $s = 100~\upmu$m.

\subsubsection{Simulation strategy}
\label{s:sim_strategy}

Due to computational limitations,
the largest particles we consider have a radius of 1~cm.
In reality, the size distribution in the parent belt extends up to planetesimals of tens to hundreds of kilometers.
To account for the fragmentation of these larger bodies,
we include
a source of dust at $ r\sub{0} $.
This artificial source term adds
particles with sizes between $ s = 1$~mm and $ s = 1$~cm,
following the power-law size distribution $ n(s) \propto s^{-3.5} $.
The size of these grains
is chosen such that the effect of radiation pressure on them is very small ($ \beta < 10^{-4} $).
Therefore, their eccentricity distribution
follows
that of the parent bodies.
We add the source particles
at eccentricities ranging from $ e = 0 $ to $ e = 0.1 $.

The artificial supply of large dust particles is balanced by the loss of material due to
blowout and sublimation.
Therefore, solving the continuity equation results in a steady-state distribution function.
We start the integration without any material in the disk
(only the artificial supply is acting), and
let the model run until the steady state is reached.
This is considered to be the case
when relative changes in the radial and size distributions
between logarithmic \mbox{(base-10)} time steps
become less than $1 \%$.\footnote{
Formally, the steady state is only reached after $ \sim 10 $~Gyr,
which is the time it takes for the largest particles we consider to move from the parent belt to the sublimation zone by \mbox{P--R} drag.
The barely bound grains that dominate the cross-section, however, already settle into
a steady state after $ \sim 10 $~Myr, which is short compared to the typical lifetime of a debris disk.}
Initially, we only consider collisions and \mbox{P--R} drag,
and the sublimation module of the code is switched off.
At this stage, particles migrate inward due to \mbox{P--R} drag until they reach the inner edge of the grid.
Once a steady state is reached, sublimation is switched on,
and we let the distribution function settle into a new steady state.
This procedure is necessary because sublimation forces the time step to become very short.
With sublimation switched on from the start,
the computation would take unnecessarily long.
Additionally, it allows us to
isolate the effect of dust pile-up
(i.e., the dust in the pile-up can be isolated by subtracting the pre-sublimation state from the final one).

\subsection{Results}
\label{s:results}

The output of each model run is a steady-state distribution of particles in the phase space of orbital elements and masses.
To analyze this output, we convert it into
a radial geometrical optical depth profile (Sect.~\ref{s:rad_distrib})
and size distributions
(in terms of cross-section density per unit size decade $A$)
at different radial locations (Sect.~\ref{s:size_distrib}).
The conversion from raw model output to these quantities is detailed in Appendix~\ref{s:app_postproc}.

\subsubsection{Radial distribution}
\label{s:rad_distrib}

\begin{figure}[!t]
  \includegraphics[width=\linewidth]{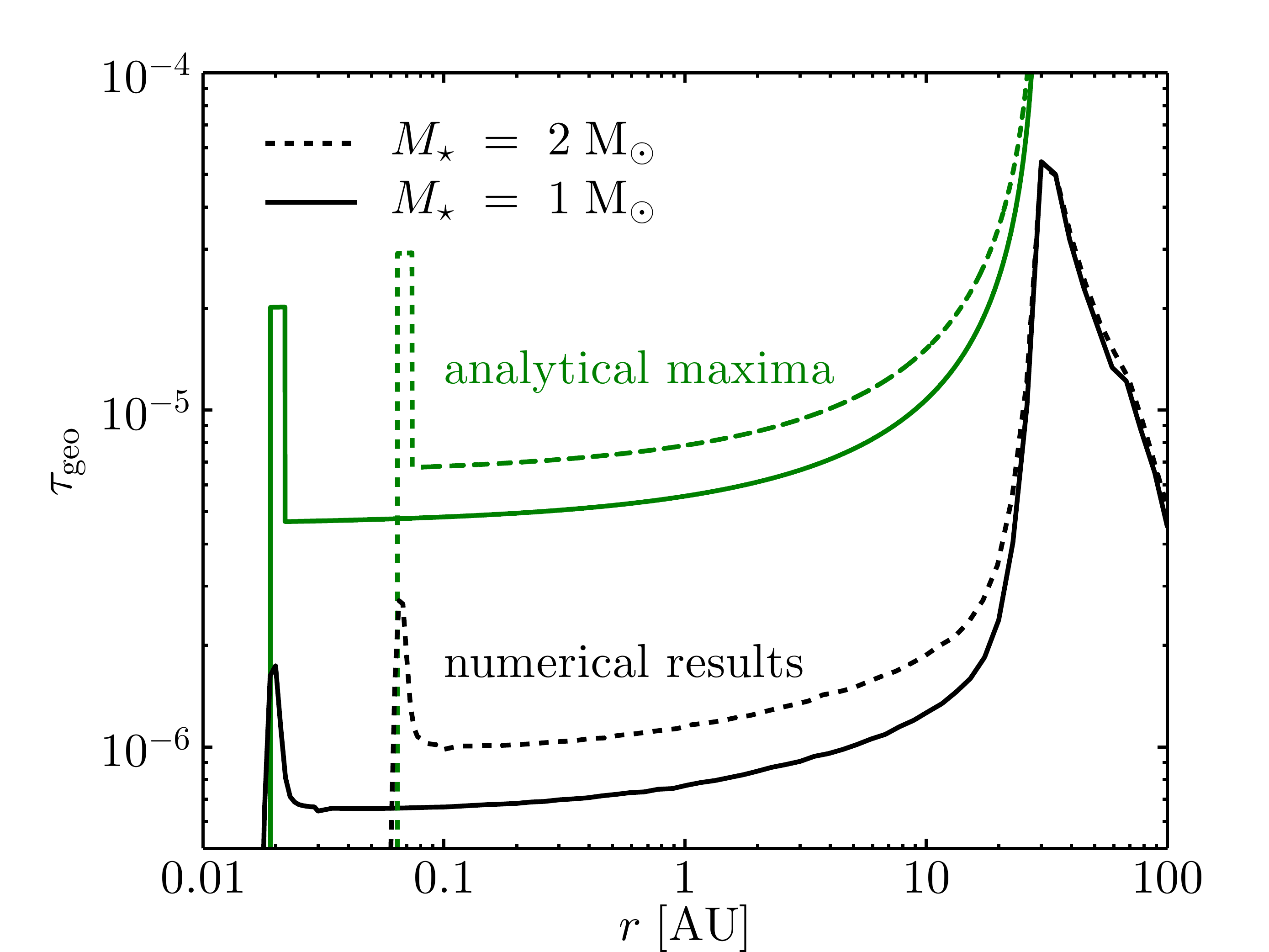}
  \caption{The geometrical optical depth profiles
  of debris disks with a dense parent belt at 30~AU,
  around stars of $ M\sub{\star} = \mathrm{1~M}\sub{\odot} $ (solid lines)
  and $ M\sub{\star} = \mathrm{2~M}\sub{\odot} $ (dashed lines).
  The black lines show the end results of the numerical simulations.
  In green are the maximum \mbox{$\tau\sub{geo}$-profiles}
  as given by the analytical model (Eq.~\ref{eq:tau_full},
  with $ \tau\sub{geo,\,base}(r) $ given by Eq.~\ref{eq:max_tau_eff},
  $ \Delta r\sub{pile} / r\sub{pile} = 0.15 $, $ r\sub{0} = \mathrm{30~AU} $, and $ \beta = 0.5 $).
  }
  \label{fig:num_tau}
\end{figure}

Figure~\ref{fig:num_tau} shows the geometrical optical depth profiles
derived from the numerical model runs,
together with the analytical maxima given by Eq.~\ref{eq:tau_full},
using a pile-up width of $ \Delta r\sub{pile} / r\sub{pile} = 0.15 $,
chosen to match the numerical profile.
Generally, there is a good correspondence between the numerical results and the analytical maxima,
but there are some important differences.
The profiles have roughly the same shape,
with a slightly steeper slope close to the source region in the numerical results.
For both cases of stellar mass, the base level of $ \tau\sub{geo} $ in the inner disk
(i.e., away from the pile-up)
is a factor of about 7 lower in the numerical profiles.
This discrepancy is a result of the the assumption in the analytical model
that all orbits are circular.
In the numerical model,
the small particles that contribute most to the cross-section
are released in the parent belt on eccentric orbits.
Hence, they suffer a higher rate of destruction by collisions.

As predicted, switching on sublimation leads to an accumulation of dust.
The pile-ups are located very close to $ r\sub{pile} $,
as determined for graphite grains,
using black-body temperatures and $Q\sub{pr} = 1$
(Eq.~\ref{eq:r_pile_approx}, which gives
$ r\sub{pile} \approx 0.019 $~AU for the $ M\sub{\star} = \mathrm{1~M}\sub{\odot} $ case,
and $ r\sub{pile} \approx 0.064 $~AU for the $ M\sub{\star} = \mathrm{2~M}\sub{\odot} $ case).
The temperature of the dust (as computed using the full optical properties)
at the inner edge of the disk, and
in the pile-up, is between 2000 and 2100~K.
In both runs, the pile-up has a geometrical optical depth enhancement factor of about $ f\sub{\tau_{geo}} \approx 3 $.
The fractional luminosities due to the pile-ups are
$ ( L\sub{D} / L\sub{\star} )\sub{pile} \approx 7.5 \times 10^{-8} $
for the $ M\sub{\star} = \mathrm{1~M}\sub{\odot} $ case,
and $ ( L\sub{D} / L\sub{\star} )\sub{pile} \approx 1.1 \times 10^{-7} $,
for the $ M\sub{\star} = \mathrm{2~M}\sub{\odot} $ case.
Both are about a factor 15 lower than the maxima given by Eq.~\ref{eq:max_fraclum_pr_r0}.
Given that the base levels of $ \tau\sub{geo} $ in the numerical profiles
are a factor of about 7 lower than the analytical maxima,
however, the discrepancy
is only about a factor of 2.
In short, the pile-up mechanism is found to be somewhat less efficient than predicted by the analytical estimates.

The outer disk ($ r > \mathrm{30~AU} $) is not the focus of this work.
Nevertheless, its radial profile is relevant,
since it reflects the status of the balance between collisions and \mbox{P--R} drag.
To first order, the geometrical optical depth profile of the outer disk
can be characterized by a power law $ \tau\sub{geo} \propto r^{-\alpha} $.
\citet{2006ApJ...648..652S} derive the theoretical values of $ \alpha = 1.5 $ and $ \alpha = 2.5 $
for collision and \mbox{P--R} drag dominated disks, respectively.
We find slopes of $ \alpha \approx 2.0 $ for both runs,
consistent with the outer slope found by \citet{2010A&A...520A..32V} for the Edgeworth-Kuiper Belt,
and interpreted as the sign of a disk that is in between drag and collision dominated.
This indicates that the density of the parent belt (characterized by $ \eta\sub{0} \sim 10 $)
is insufficient to make the outer disk completely collision dominated.

\subsubsection{Size distribution}
\label{s:size_distrib}

\begin{figure}[!t]
  \includegraphics[width=\linewidth]{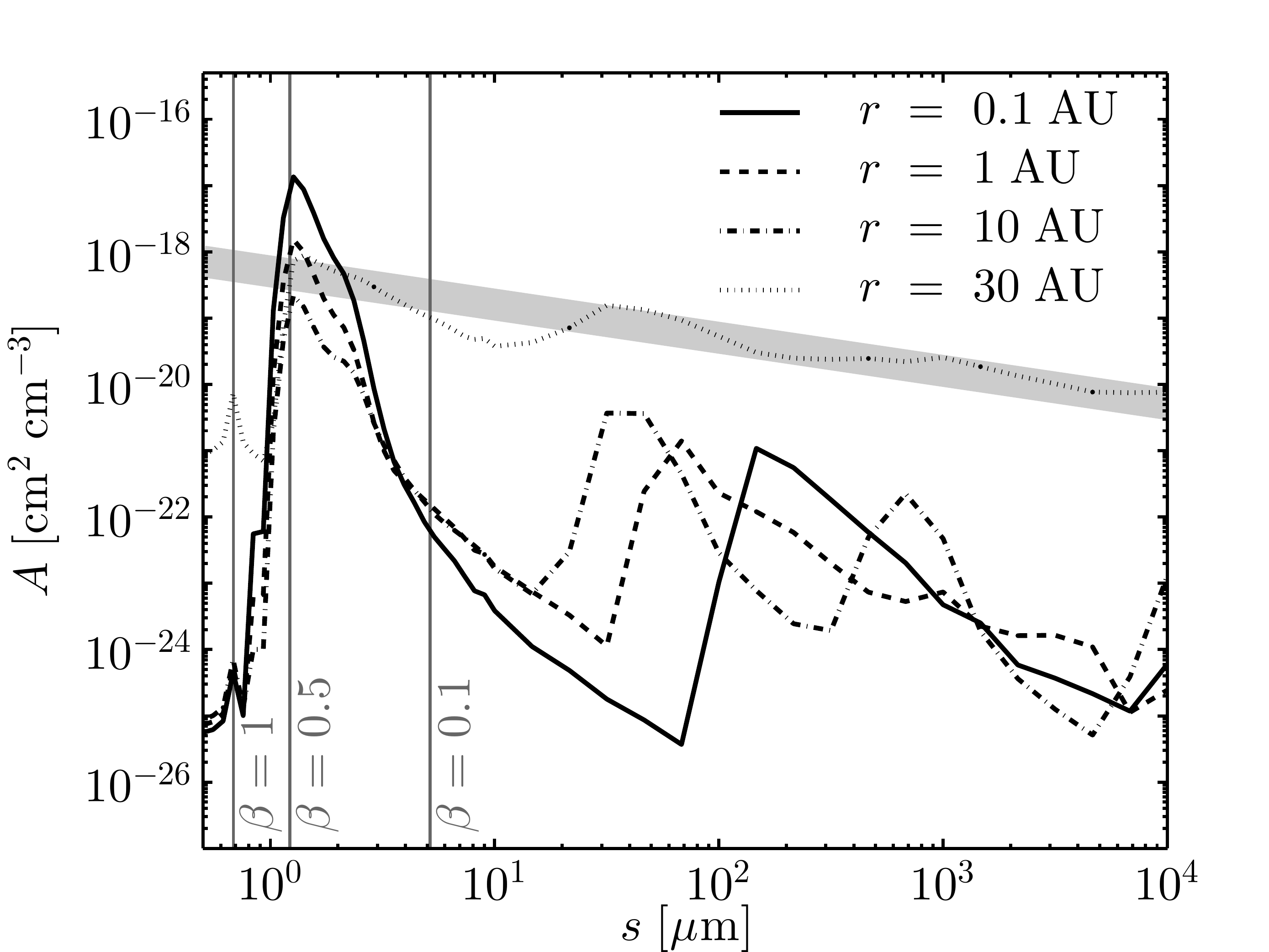}
  \caption{Size distributions at different radial distances
  for the $ M\sub{\star} = \mathrm{1~M}\sub{\odot} $ run,
  before switching on sublimation.
  The quantity on the vertical axis, $A$, is the cross-section density per unit size decade
  (see Appendix~\ref{s:app_postproc_quantities}),
  which is horizontal if all sizes contribute equally to the cross-section.
  The grey band indicates the slope of a size distribution that follows
  the classical \citet{1969JGR....74.2531D} power law ($ n(s) \propto s^{-3.5}$, and hence $ A \propto s^{-0.5} $).
  The vertical lines mark the particle sizes
  corresponding to
  relevant values of $\beta$.
  }
  \label{fig:size_prtail}
\end{figure}

The size distribution results of the two runs are similar in many ways,
so we only discuss the $ M\sub{\star} = \mathrm{1~M}\sub{\odot} $ run here.
Figure~\ref{fig:size_prtail} shows
how the size distribution changes with radial distance,
only considering \mbox{P--R} drag and collisions
(i.e., before sublimation is switched on in the model).
In the parent belt ($ r = 30 $~AU),
it follows the classical
\citet{1969JGR....74.2531D} power law ($ n(s) \propto s^{-3.5}$,
valid for an infinite collisional cascade with self-similar collisions),
from the blowout radius upwards.
Particles with $ \beta > 0.5 $ are depleted by about three orders of magnitude
in terms of collective cross-section.
Superimposed on the power law is a well-known wave pattern
related to the discontinuity in the size distribution at the blowout size
\citep[see, e.g.,][]{1994P&SS...42.1079C,1997Icar..130..140D,2012ApJ...754...74G}.
The first bump (i.e., the one at $ \beta \approx 0.5 $) in the size distribution at $r = 30$~AU
does not extend far above the power-law prediction.
The reason for this may be that particles with $ \beta \gtrsim 0.1 $ experience more destructive collisions,
because their eccentricities are significantly higher than those of the parent bodies,
which are distributed over the range $ 0 < e < 0.1 $
(Sect.~\ref{s:sim_strategy}, see also Eq.~\ref{eq:e_beta}, and cf. Fig.~5 of \citealt{2006A&A...455..509K}).

Inwards from the parent belt, the size distribution seems to become steeper,
which is expected from the dependence of the radial profile on $\beta $ (Eq.~\ref{eq:max_tau_eff}).
However, this effect is difficult to isolate, because the profile is distorted by the wave pattern,
which increases in amplitude and ``wavelength"
with decreasing radial distance \citep[cf. Fig.~7 of][]{2006A&A...455..509K}.
The prominent wave pattern 
indicates that collisions are still important for larger particles in the inner disk
(while \mbox{P--R} drag dominates for $\beta \approx 0.5$ particles there, see Fig.~\ref{fig:timescales}).
In the innermost parts of the disk ($ r \lesssim 1 $~AU),
particles with $ \beta \approx 0.5 $ clearly dominate the cross-section,
contributing at least three orders of magnitude more than any other size.
At $ r = 0.1 $~AU, the local slope of the size distribution
between the blowout size and $ s \approx 10~\upmu$m
is approximately $ A \propto s^{-7.5} $, equivalent to $ n(s) \propto s^{-10.5}$.
This means that not only the cross-section,
but also the mass is dominated by barely bound grains in the innermost parts of the disk.
Interestingly, such steep size distributions are also
invoked to explain NIR interferometric observations of
hot exozodiacal dust \citep{2011A&A...534A...5D,2013ApJ...763..119M,2013A&A...555A.146L}.
The drop in the size distribution from $ \beta \approx 0.5 $ to $ \beta \approx 1 $
is also much more pronounced in the inner disk compared to the parent belt.

\begin{figure}[!t]
  \includegraphics[width=\linewidth]{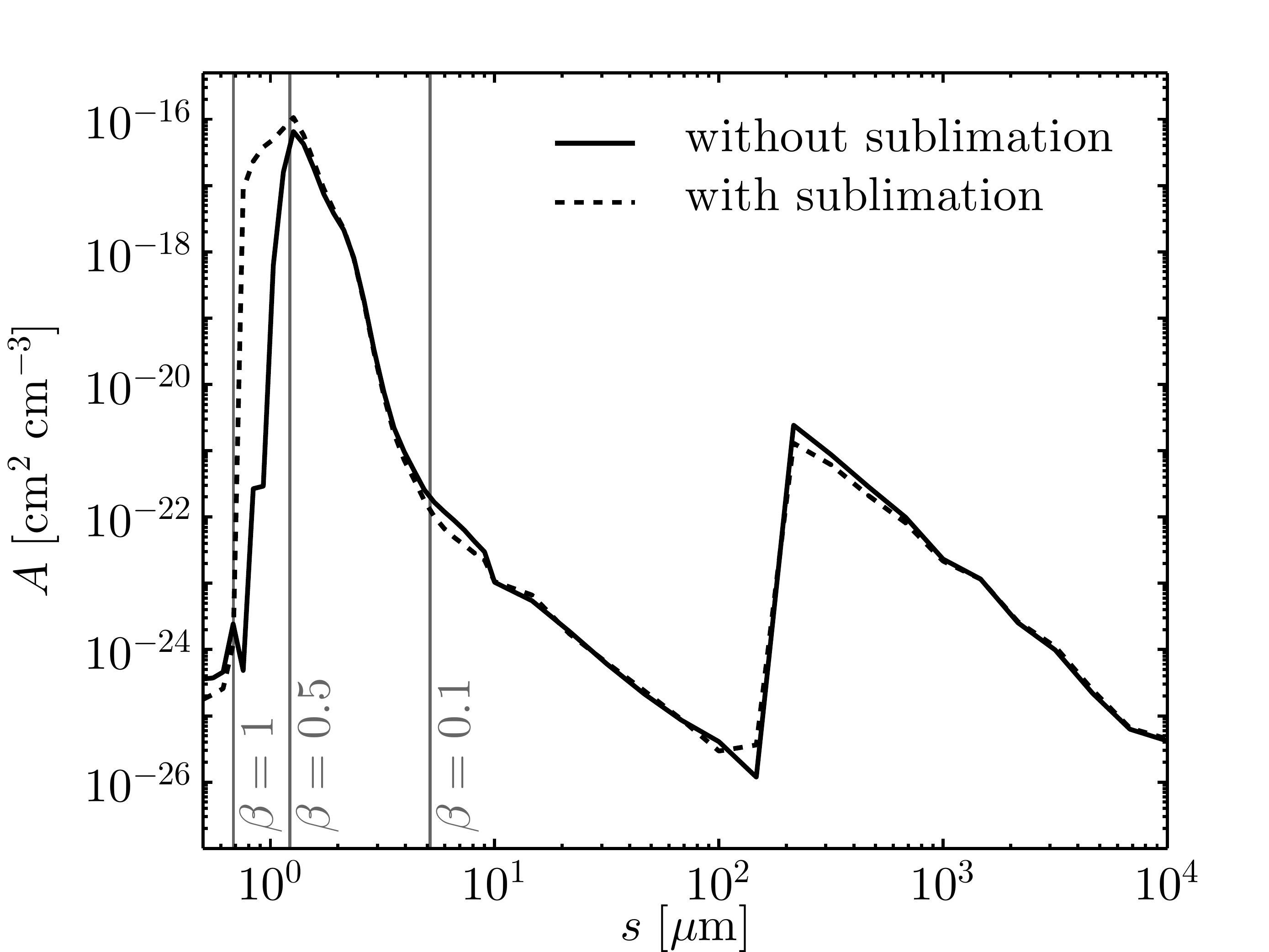}
  \caption{The size distribution at $ r = 0.02$~AU (the location of pile-up)
  for the $ M\sub{\star} = \mathrm{1~M}\sub{\odot} $ run,
  before and after sublimation is switched on.
  }
  \label{fig:size_pileup}
\end{figure}

Sublimation only has a significant effect on the size distribution
around $ r \approx r\sub{pile} $ (i.e., in the pile-up).
Figure~\ref{fig:size_pileup} shows
the size distribution in the pile-up,
before and after sublimation is switched on.
It clearly demonstrates that
sublimation enhances the density of particles with $ 0.5 \lesssim \beta < 1 $ around $ r = r\sub{pile} $.
This indicates that the pile-up consists mostly of particles that
started with $ \beta \approx 0.5 $ before active sublimation,
and lost mass due to sublimation, increasing their $ \beta $.
The rest of the size distribution does not change significantly.
Size distributions at larger radial distances are largely unaffected by sublimation.

\section{Comparison with observations}
\label{s:seds}

\begin{figure*}[!t]
  \includegraphics[width=\linewidth]{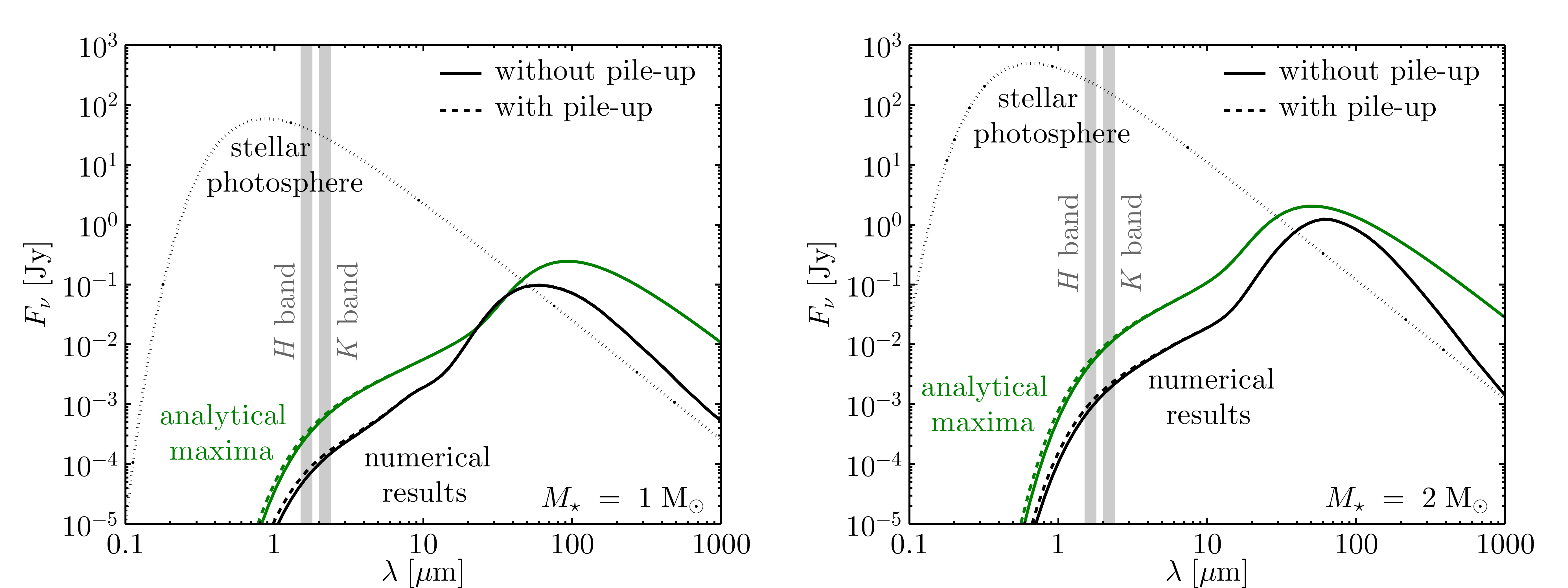} 
  \caption{The SEDs
  of stars and their debris disks at a distance of 10~pc.
  The disk SED is shown for different dust distributions,
  with the numerical results in black,
  and the analytical distributions in green.
  Dust distributions including the pile-up of dust in the sublimation zone are shown with dashed lines,
  and distributions in which the pile-up is excluded are shown with solid lines,
  but the spectra largely overlap.
  Disk SEDs are calculated according to Eqs.~\ref{eq:sed_dust} and \ref{eq:sed_dust_num}.
  The stellar photosphere is indicated by a dotted Planck curve.
  The vertical grey areas mark the NIR \textit{H} and \textit{K} bands
  in which hot exozodiacal dust is observed.
  }
  \label{fig:seds}
\end{figure*}

In order to assess whether 
the pile-up effect can explain the observed NIR excess,
we compute the spectral energy distributions (SEDs)
of the dust distributions found in the previous sections.
We only calculate the emission spectrum of the dust,
and ignore the (viewing angle dependent) contribution of scattered light,
since thermal emission was found to dominate scattered light at the wavelengths
in which hot exozodiacal dust is detected \citep{2006A&A...452..237A}.
For the analytical dust distributions, we assume the dust has a black-body temperature (Eq.~\ref{eq:temp_bb}).
The flux density of the dust,
expressed 
as a function of wavelength $\lambda$, can then be computed as
\begin{equation}
  \label{eq:sed_dust}
  F_{\nu,\;\!\mathrm{D}}(\lambda) =
  \frac{ 2 \pi \lambda^2 }{ c d^2 } \int r \, \tau\sub{geo}(r) B_\lambda(T\sub{bb}) \, \dif r,
\end{equation}
where $d$ is the distance to the source, and $ B_\lambda(T) $ is the Planck function.
For the numerically determined dust distributions,
we use realistic dust temperatures and optical properties (see Sect.~\ref{s:setup_material}),
which leads to
\begin{equation}
  \label{eq:sed_dust_num}
  F_{\nu,\;\!\mathrm{D}}(\lambda) =
  \frac{ 2 \pi \lambda^2 }{ c d^2 }
  \int\limits_{s} \! \! \! \int\limits_{r} r Q\sub{abs}(s) \, \tau\sub{geo}(s,r) B_\lambda[T(s,r)] \, \dif s \, \dif r.
\end{equation}
Here, $ Q\sub{abs} $ is the absorption efficiency of the grains (equal to their emission efficiency)
and $ \tau\sub{geo}(s,r) $ is the geometrical optical depth profile of dust with grain radius $s$.

Figure~\ref{fig:seds} shows the debris disk spectra
corresponding to dust distributions found from
the numerical modeling,
as well as the analytical maximal dust profiles,
together with the stellar spectra.
The analytical maxima are computed from Eq.~\ref{eq:tau_full},
with $ \tau\sub{geo,\,base}(r) $ given by Eq.~\ref{eq:max_tau_eff}, $ r\sub{0} = 30$~AU, $ \beta = 0.5 $,
and all the pile-up material at $ r\sub{pile} $ (i.e., $ \Delta r\sub{pile} / r\sub{pile} \ll 1 $).\footnote{
Note that our approach in computing the SED of a dust profile with pile-up
is different from that of \citet{2011EP&S...63.1067K},
who use $ \tau\sub{geo,\,pile} $ and $ \Delta r\sub{pile} $ as independent input variables.
In our estimate,
the total amount of dust is fixed,
and placing it all at $ r\sub{pile} $
is the most optimistic configuration for detecting the pile-up.
}
Singularities at $ r = r\sub{0} $ are removed by
imposing the condition $ \tau\sub{geo} \leq 0.01 $,
which corresponds to $ \eta\sub{0} \approx 2200 $.
The numerical dust distributions without pile-up were created by subtracting
the isolated pile-up dust from the final profile (see Sect.~\ref{s:sim_strategy}).
The stellar spectra are given by black-body curves, with
$ L\sub{\star} = \mathrm{1~L}\sub{\odot} $, $ T\sub{\star} = 5780$~K
for the $ M\sub{\star} = \mathrm{1~M}\sub{\odot} $ star
and $ L\sub{\star} = \mathrm{11.31~L}\sub{\odot} $, $ T\sub{\star} = 7730$~K
for the $ M\sub{\star} = \mathrm{2~M}\sub{\odot} $ star,
following the main-sequence relations $ L\sub{\star} \propto {M\sub{\star}}^{3.5} $
and $ T\sub{\star} \propto {L\sub{\star}}^{0.12} $ \citep{1976asqu.book.....A}.
The distance is arbitrarily set to 10~pc.

The synthetic spectra are similar to those described by \citet{2005A&A...433.1007W},
but are truncated
at short wavelengths because there is no material with a higher temperature than the sublimation temperature.
Apart from an overall shift in flux, the differences between the analytical and numerical SEDs are minor.
In the solar-mass star run, the peak of emission of the numerical result
is shifted towards shorter wavelengths with respect to the analytical SED.
This is because the grains that dominate the emission in the parent belt
are significantly hotter than the black-body temperature.

Only considering thermal emission,
the NIR flux originates almost exclusively from the inner 1~AU of the debris disk.
The pile-up does not add a significant amount of flux.
The theoretical SEDs display
NIR flux ratios between the disk and star of about $ F_{\nu,\;\!\mathrm{D}} / F_{\nu,\;\!\star} \sim 10^{-4}$.
This is much smaller than the observed flux ratios that indicate hot exozodiacal dust,
which are typically of the order of 1\% (see Tbl.~\ref{tbl:obs}).
Furthermore, the NIR radiation is accompanied by mid-IR flux at a comparable level,
which is incompatible with observed excess spectra \citep[e.g.,][]{2009ApJ...691.1896A}.
The pile-up does not contain enough material to
create a bump in thermal flux in the NIR.

\begin{figure}[!t]
  \includegraphics[width=\linewidth]{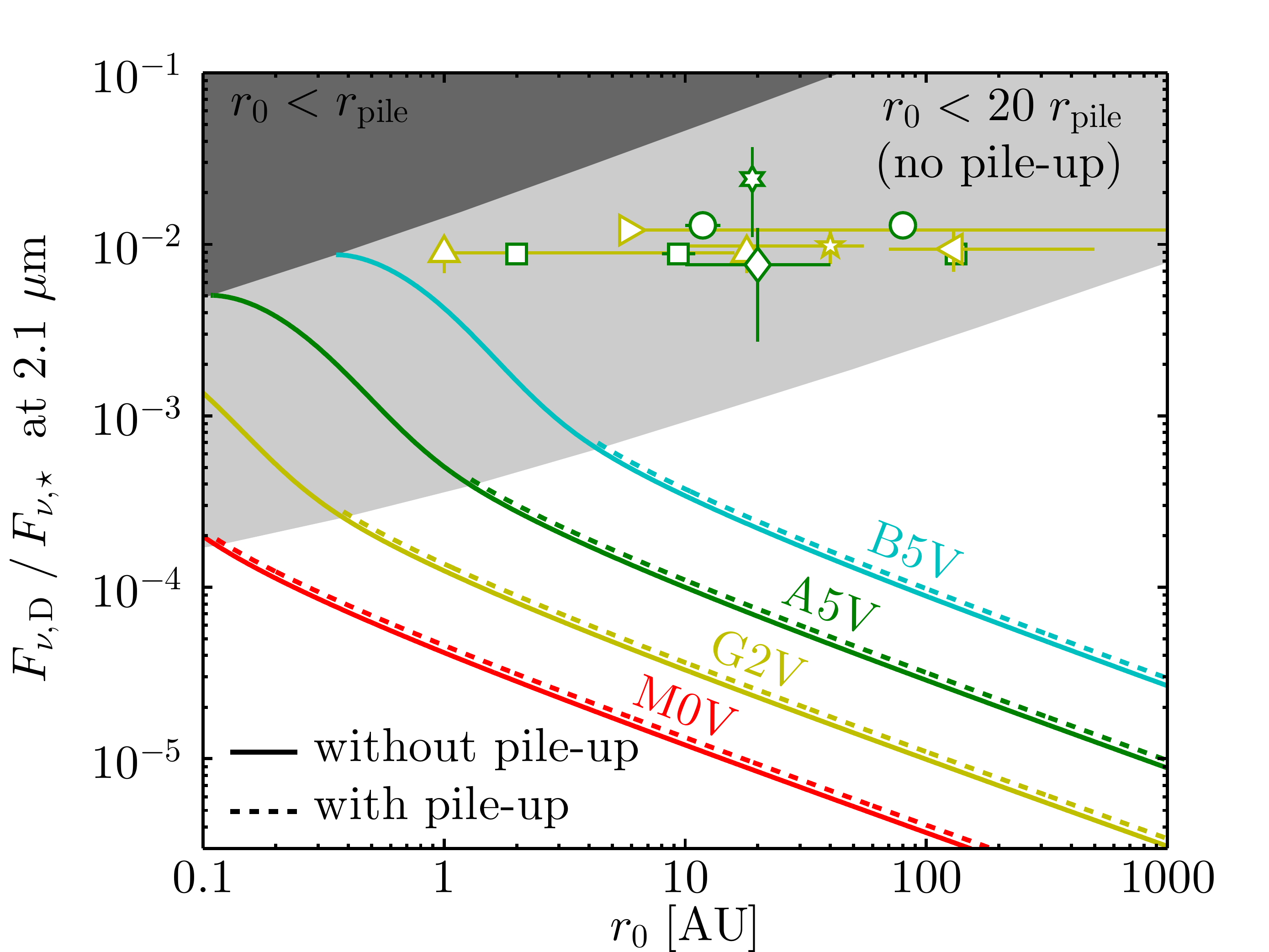}
  \caption{Flux ratio between the disk and the star at $2.1~\upmu$m \mbox{(\textit{K} band)},
  as function of parent belt radius $r_0$.
  Solid lines indicate the flux ratio due to the maximum amount of material moved in by \mbox{P--R} drag,
  truncated at $r = r\sub{pile}$.
  Ratios including the small effect due to pile-up are shown with dashed lines.
  Different colors are used for different stellar types.
  The light shaded area marks the region in parameter space
  where no significant pile-up is expected to occur
  because the orbits of small dust grains cannot circularize sufficiently (see Eq.~\ref{eq:r_pro_lim}).
  The dark shaded area marks where the parent belt $r_0$ is
  closer than the pile-up radius $r\sub{pile}$.
  The observed \mbox{\textit{K}-band} excess fluxes
  of main-sequence stars from Tbl.~\ref{tbl:obs}
  are shown at their respective estimated parent belt distances,
  with a different symbol for each star, and coloring according to the star's spectral type:
  Vega (the CHARA/FLUOR measurement): circles;
  $\upbeta$~Leo: hexagon;
  Fomalhaut: squares;
  $\upbeta$~Pic: diamond;
  $\upeta$~Lep: upward pointing triangles;
  110~Her: left-pointing triangle,
  10~Tau: right-pointing triangle;
  $\uptau$~Cet: pentagon.
  }
  \label{fig:flux_ratio}
\end{figure}

In order to generalize the above results,
we investigate how the NIR flux ratio depends on parent belt distance and stellar type using the analytical model.
In Fig.~\ref{fig:flux_ratio}, we show the analytical maximum NIR flux ratios for four different stellar types.
As in the above analysis, the disk flux is calculated from Eq.~\ref{eq:sed_dust},
using the analytical maximum dust distribution,
and the stars are approximated by black bodies.
The M0V, G2V, A5V, and B5V stars correspond to
$M\sub{\star} = 0.5$, 1, 2, and 4~M$\sub{\odot}$, respectively,
with $ L\sub{\star} \propto {M\sub{\star}}^{3.5} $
and $ T\sub{\star} \propto {L\sub{\star}}^{0.12} $ \citep{1976asqu.book.....A}.
Comparing the analytical maximum NIR flux ratios to the observed ones
demonstrates that
\mbox{P--R} drag does not provide enough material to the inner disk to explain the observations.

\section{Discussion}
\label{s:discussion}

\subsection{The size distribution in the inner disk}
\label{s:discuss_size_distrib}

The size distribution in the inner parts of dense debris disks (i.e., inwards of the parent belt)
has not been studied many times before.
\citet{2012A&A...540A.125A} analytically derive
a power-law distribution of $ n(s) \propto s^{-3} $,
but ignore collisions in the inner disk.
In a detailed modeling study of the debris disk around $\upvarepsilon$~Eri,
\citet{2011A&A...527A..57R} find a flat profile
($ n(s) \propto s^{-3} $, flat in terms of $A$) for small sizes,
and a steeper one ($ n(s) \propto s^{-3.7} $) for larger sizes.
However, this system is a special case,
because $\upvarepsilon$~Eri is not luminous enough to blow small dust grains out of the system.

With our numerical model, we find a size distribution that deviates significantly from a power law.
Specifically, in the innermost regions of the disk ($r \lesssim 1$~AU),
the cross-section is completely dominated by barely bound ($\beta \approx 0.5$) particles.
This is the result of two effects:
(1)~Since the efficiency of \mbox{P--R} drag depends on $\beta$,
larger particles tend to stay closer to the parent belt (apparent from Eq.~\ref{eq:max_tau_eff}).
(2)~Due to the high relative velocities in the inner disk,\footnote{
In debris disks relative velocities are mostly due to the eccentricities and mutual inclinations of particle orbits.
While orbital eccentricities are diminished in the inner disk by \mbox{P--R} drag induced circularization,
inclinations are simply inherited from the parent belt.
This is handled correctly in our code, since orbital inclinations are parameterized
by the (constant) opening angle of the disk.}
the wave in the size distribution is extremely strong.
As a consequence, the single size assumption used in our analytical model is a good approximation.

Erosive (cratering) collisions affect the size distribution \citep{2007A&A...472..169T}.
We expect that including this type of collisions in our model would result
in a less wavy size distribution,
possibly eliminating the second bump in the size distribution \citep[see Sect.~4.3.4 of][]{2008PhDTLohne}.
This would mean that $ \beta \approx 0.5 $ particles dominate the cross-section even more
than suggested by Fig.~\ref{fig:size_prtail}.
Since erosive collisions present an additional mechanism of destroying dust,
we expect that including them into our model would only lower the level of dust in the inner disk,
and therefore does not change the conclusions of this work.
Erosive collisions do affect
the timescale on which debris disks evolve,
but since we compute steady-state dust distributions,
this does not have an impact on our results.

It may be surprising that a large amount of particles are present in the pile-up with $ 0.5 \lesssim \beta < 1 $.
Usually (as in Sect.~\ref{s:analytic}), particles with $ \beta > 0.5 $ are assumed to be absent,
because they are blown out of the system as soon as they are released from parent bodies on circular orbits.
Parent bodies on eccentric orbits can give rise to a population of bound grains with $ \beta > 0.5 $,
but the parent body orbits  in our model runs
are not eccentric enough to produce bound dust grains with $ \beta $ values close to unity.
The origin of this high $ \beta $ population is the sublimation of barely bound $ \beta \approx 0.5 $ grains.
As these particles migrate inward from the parent belt, their orbits are circularized by \mbox{P--R} drag.
When they arrive in the sublimation zone,
their orbital eccentricities are as low as $ e \approx 10^{-4} $.
\citet{2009Icar..201..395K} find that the subsequent evolution of the eccentricities
during the active sublimation phase can be described by
\begin{equation}
  \label{eq:e_pile}
  e = \left( \frac{ 1 - \beta_1 }{ 1 - \beta } \right)^\kappa e_1,
\end{equation}
where subscript 1 denotes quantities at the start of substantial sublimation,
and the exponent $ \kappa $ can be treated as a constant,
which depends mostly on the optical and sublimation properties of the material under consideration.
For spherical, black-body graphite particles we find $ \kappa \approx 10 $.
This shows that particles starting with $ \beta_1 \approx 0.5 $, and $ e_1 \approx 10^{-4} $
will only become unbound ($ e \geq 1 $) when they reach $ \beta \gtrsim 0.8 $
(corresponding to $s \approx 0.8~\upmu$m).

\subsection{Stellar wind drag}
\label{s:discuss_wind_drag}

Neither our analytical, nor our numerical model includes the effects of stellar wind.
For stars with a strong stellar wind and/or a low luminosity,
the stellar wind equivalent of \mbox{P--R} drag shortens the inward migration timescale,
and hence increases the maximum geometrical optical depth profile.
Since stellar wind drag works in the same way as \mbox{P--R} drag,
its effect can be accounted for by replacing $\beta$ with
\citep{1979Icar...40....1B,2004A&A...424L..13M,2006A&A...452..701M}
\begin{subequations}
  \label{eq:beta_wind}
  \begin{align}
    \beta\sub{PR} & =  \beta ( 1 + \gamma ), \\
    \label{eq:beta_wind_gamma}
    \gamma & = \frac{ \dot{ M }\sub{\star} c^2 }{ L\sub{\star} } \frac{ Q\sub{sw} }{ Q\sub{pr} },
  \end{align}
\end{subequations}
where $ \dot{ M }\sub{\star} $ is the stellar mass loss rate by stellar wind,
and $ Q\sub{sw} $ is the stellar wind momentum transfer efficiency.
For carbonaceous particles orbiting the Sun,
\citet{2004A&A...424L..13M,2006A&A...452..701M} find $ \gamma \sim 1 $.

\citet{2011EP&S...63.1067K} argue that the pile-up scenario could work for Vega
if the disk is drag dominated,
which requires $\gamma \approx 300$.
While this value is consistent with the upper limit on the mass-loss rate of Vega from radio-continuum observations
\citep[$ \dot{ M }\sub{\star} \lesssim 10^{-10}$~M$\sub{\odot}$~yr$^{-1}$;][]{1985ApJ...294..646H},
stellar wind models predict much lower mass-loss rates for main-sequence \mbox{A-type} stars
\citep[$ \dot{ M }\sub{\star} \lesssim 10^{-16}$~M$\sub{\odot}$~yr$^{-1}$;][]{1995A&A...301..823B}.
This theoretical mass-loss rate,
together with $ L\sub{\star} \approx 10~\mathrm{L}\sub{\odot}$ and $ Q\sub{sw} \approx Q\sub{pr} $,
gives a ratio between stellar wind drag and \mbox{P--R} drag of $\gamma \lesssim 10^{-4}$.
Hence, we conclude that it is unlikely that stellar wind drag has a significant effect on
debris disks around main-sequence \mbox{A-type} stars.

\subsection{Other explanations for hot exozodiacal dust}
\label{s:discuss_general}

We find that \mbox{P--R} drag does not provide enough material to the innermost parts of the disk
to explain the interferometric detections of NIR excess.
There are two additional problems with this scenario,
that also serve as clues for solving the hot exozodiacal dust mystery.
(1)~Hot exozodiacal dust is thought to consist mostly of blowout grains with sizes around \mbox{0.01--0.1~$\upmu$m}
\citep{2009ApJ...691.1896A,2011A&A...534A...5D,2013ApJ...763..119M,2013A&A...555A.146L},
while \mbox{P--R} drag only transports bound grains to the sublimation zone,
the smallest of which have sizes of about 1~$\upmu$m.
(2)~The dust distribution resulting from the balance between \mbox{P--R} drag and collisions yields
an SED with a positive slope in the infrared domain,
while observations
find negative slopes
\citep[e.g.,][]{2009ApJ...691.1896A,2012A&A...540A.125A},
and the pile-up of dust is too inefficient to have an effect on the slope of the SED.\footnote{
For the $ M\sub{\star} = \mathrm{1~M}\sub{\odot} $, $r_0 = 30$~AU case,
the pile-up efficiency should be at least $\sim$100 times higher to make the slope of the SED negative.}
For these reasons, the origin of hot exozodiacal dust
must involve mechanisms that we did not consider in this work.

Our treatment of sublimation assumes spherical dust grains
that sublimate uniformly
(i.e., layers of material are removed one by one).
In reality, dust grains may be aggregates that fall apart during sublimation,
abruptly increasing the collective cross-section of the material.
To investigate this scenario,
the steady-state amount of material can be estimated by multiplying
the maximum \mbox{P--R} drag inward mass flux (Eq.~\ref{eq:max_mass_flux_r0})
with an estimate of the lifetime of the fragments.
We performed this analysis for Fomalhaut,
and find that \mbox{P--R} drag from a parent belt at 2~AU still does not supply enough material,
unless the lifetime of the
fragments is significantly longer than what can be expected from sublimation and blowout
\citep{2013A&A...555A.146L}.

Two mechanisms have recently been proposed that can lead to an extended lifetime
for small exozodiacal dust particles.
\citet{2013A&A...555A.146L} investigate the impediment of blowout
due to the presence of gas for the hot dust around Fomalhaut,
but find that this requires unrealistically high gas densities.
\citet{2013ApJ...763..118S} propose that charged nanograins can remain trapped in the magnetic field of the star,
and qualitatively show that this
may help explain the NIR excess of Vega.

Another mechanism for the inward transport of material from a cold outer belt is
the inward scattering of material by planets.
\citet{2012A&A...548A.104B} investigated this scenario,
and found that it is marginally
capable of providing mass influxes compatible with observations.
However, this requires relatively contrived planetary system architectures,
consisting of closely packed chains of low mass planets.
Furthermore, the scenario was investigated for the inward scattering down to 1~AU,
and reaching the sublimation zone is likely to be less efficient.

\section{Conclusions}
\label{s:conclusions}

In this work, we
investigated 
hot dust in the inner regions of debris disks,
whose presence is
suggested by interferometrically resolved excess NIR emission
observed in some debris disk systems (Tbl.~\ref{tbl:obs}).
We tested whether the hot dust can be supplied by \mbox{P--R} drag from a distant parent belt,
and whether the pile-up of dust in the sublimation zone still occurs if collisions are considered.
Our main conclusions are as follows:
\begin{enumerate}
\item As predicted by \cite{2005A&A...433.1007W}, \mbox{P--R} drag always brings a small amount of dust
from an outer debris belt into the sublimation zone.
The maximum geometrical optical depth that can be reached in the innermost parts of the disk
depends on the mass of the central star and distance to the parent belt (Fig.~\ref{fig:max_tau}).
When the production of dust is treated self-consistently,
this maximum is found to be a factor of about 7 lower
than the analytical estimate (Fig.~\ref{fig:num_tau}).
This is because small dust particles, which are dragged inwards efficiently by radiation forces,
are also put on highly eccentric orbits by those radiation forces,
and therefore suffer more collisional destruction.
\item Dust that reaches the sublimation zone
produces some NIR emission,
but this excess flux
is insufficient to explain the interferometric observation.
While the observed excess ratios are of the order of $\sim$10$^{-2}$,
the maximum flux ratio due to material supplied by \mbox{P--R} drag
is $\lesssim$10$^{-3}$ for \mbox{A-type} stars with parent belts at $\gtrsim$1~AU (Fig.~\ref{fig:flux_ratio}).
\item The pile-up of dust due to the interplay of \mbox{P--R} drag and sublimation
still occurs when collisions are considered (Fig.~\ref{fig:num_tau}),
as long as the parent belt from which the dust originates is distant enough
to allow for sufficient circularization of the orbits,
and the central star is luminous enough to blow small dust grains out of the system.
Collisions do not interfere with the pile-up process,
since in the inner disk, the collisional timescale is longer than the \mbox{P--R} drag timescale
for the barely bound grains that are the most important for the pile-up.
The fractional luminosity provided by dust in the pile-up is relatively small,
so the pile-up does not influence the disk SED significantly (Fig.~\ref{fig:seds}).
\item In the inner parts of dense debris disks,
the cross section is clearly dominated by barely bound ($\beta \approx 0.5$) grains,
and the size distribution 
features a prominent wave pattern, related to the discontinuity in the size distribution at the blowout size
(Fig.~\ref{fig:size_prtail}).
In the pile-up, there is an enhancement of particles
with $0.5 \lesssim \beta < 1$ (Fig.~\ref{fig:size_pileup}).
These particles are still bound,
because of their almost circular orbits at the start of substantial sublimation.
\end{enumerate}

\begin{acknowledgements}
We thank the anonymous referee for a very thorough and detailed report,
that helped us to improve the quality of this paper.
\end{acknowledgements}

\bibliographystyle{aa}
\bibliography{bib_ads.bib}

\begin{appendix}

\section{Numerical techniques}
\label{s:app_num}

Our model is computationally very demanding, primarily due to the large amount of calculations needed for collisions,
which scales with the
number of phase-space
bins cubed.
To ensure that the model can produce results in a reasonable amount of time,
an effort was made to reduce the amount of calculations necessary per run.
To achieve this, we used an integration technique that allows large time step sizes,
which reduces the number of steps needed per run (App.~\ref{s:app_integrate}).
Furthermore, we calculated as many time-independent numerical factors as possible before the actual integration,
so these factors do not have to be determined at every time step (App.~\ref{s:app_precalc}).

\subsection{Integration method}
\label{s:app_integrate}

After discretization, the distribution function $n( m, \vec{k}, t )$ is replaced by the vector of population levels $ \vec{n}$,
and Eq.~\ref{eq:master} can be treated as the system of linear equations
\begin{equation}
  \label{eq:sys_lin}
  \frac{ \dif \vec{n} }{ \dif t } = \vec{A} \vec{n} + \vec{b},
\end{equation}
where $ \vec{A} $ is a matrix of coefficients, and $ \vec{b} $ is a constant vector
containing the artificial source terms that replenish dust in the parent belt.\footnote{
Collision rates are proportional to target and projectile densities,
so the entries of matrix $ \vec{A} $ contain terms with elements of $ \vec{n} $.
Because target and projectile bin can be the same,
strictly the system of equations is non-linear.}
This system of equations
suffers from stiffness:
the population levels of some bins change very rapidly compared to others,
mainly due to large differences in collisional timescales,
and the time step size is therefore determined by the stability of the integration method rather than its accuracy.
When using standard explicit integrators,
this leads to an impractically small step size that prevents long integrations.

Following \citet{2008PhDTLohne}, we resolve the stiffness by
writing the differentials as
\begin{equation}
  \label{eq:split}
  \frac{ \dif \vec{n} }{ \dif t }
  = \vec{ A' } \, \vec{n} + \left. \frac{ \dif \vec{n} }{ \dif t } \right|\sub{const},
\end{equation}
where $ \vec{A'} $ is a diagonal matrix, containing only the diagonal elements of $ \vec{A} $,
while the terms marked ``const'' contain the off-diagonal parts of $ \vec{A} $, as well as the constant terms $ \vec{b} $.
Our integration scheme
for the \mbox{$j \; \! $\textsuperscript{th}} component of $ \vec{n} $, for timestep $m+1$,
reads
\begin{equation}
  \label{eq:exp_eul}
  n_{ j, \; \! m + 1 }
    = n_{ j, \; \! m } \exp \left( A'_{ jk, \; \! m } \Delta t \right)
      + \frac{ \dot{n}_{ j, \; \! m }| \sub{const} }{ A'_{ jk, \; \! m } }
        \left[ \exp \left( A'_{ jk, \; \! m } \Delta t \right) - 1 \right],
\end{equation}
where $ \Delta t $ denotes the time step size.\footnote{
This corresponds to Eq.~3.107 of \citet{2008PhDTLohne},
corrected for  a typographical error.}
This scheme is known as the exponential Euler method
(see \citealt{ANU:7701740} for an introductory review).
It is suitable for time integration of semi-linear problems,
which consist of a stiff linear part, and a non-stiff non-linear part.
In short, the strategy is to solve the linear part exactly,
and approximate the non-linear part using an explicit integration scheme.

The time step size $ \Delta t $ is determined
dynamically
from the condition that population levels can never become negative.
We use a scheme similar to that of \citet{2005Icar..174..105K},
but adapted for the exponential Euler method.

\subsection{Precalculation}
\label{s:app_precalc}

Only considering collisions (i.e., ignoring diffusion terms),
the evolution of
the \mbox{$j \; \! $\textsuperscript{th}} component of $ \vec{n} $
can be written as
\citep{2008PhDTLohne}
\begin{equation}
  \label{eq:precalc}
  \dot{ n }_{ j } = \displaystyle\sum\limits_{tp} B_{jtp} n_{t} n_{p}.
\end{equation}
Here, $t$ and $p$ are the target and projectile bin indices, respectively,
and $ B_{jtp} $ is an entry in the time-independent tensor of collisional coefficients $ \tens{B} $.
Specifically, coefficient $ B_{jtp} $ is the
rate at which the population level of bin $j$ changes,
per particle in bin $t$, per particle in bin $p$,
combining
all considered relative orbit orientations.
If $ j = t$ or $ j = p $, this is the loss rate of the target or projectile bin, respectively, due to collisional destruction
(assuming that the mass grid resolution is high enough that fragments do not end up in their parent bin).
Otherwise, it is a rate at which fragments are created.

For the phase-space grids we use, the entire tensor $ \tens{B} $ is too large to be stored in memory.
However, it is very sparse, because
(1)~not all orbits that are part of the phase-space grid have mutual overlap,
(2)~overlap may occur for a limited range in relative orbit orientation,
(3)~impact velocities are not always high enough to cause catastrophic collisions, and
(4)~only a fraction all possible masses are created as fragments.

By only storing the non-zero entries of $ \tens{B} $,
it becomes possible to keep it in memory.
We store the source and sink terms separately.
For the source terms, which are by far the most memory intensive, each non-zero entry requires
(1)~the index of the target bin,
(2)~the index of the projectile bin, 
(3)~the index of the bin fragment bin, and
(4)~the rate at which particles are created in the fragment bin,
per particle in the target bin, per particle in the projectile bin.
By further manual compression, only (3) and (4) need to be stored for each entry separately.
The target bin index~(1) only needs to be stored once for each possible target bin,
along with the number of possible projectile bins for that target bin.
Then, the projectile bin index~(2) only needs to be stored once for each possible collisional pair of bins,
along with the number of possible fragment bins for that pair.
A similar scheme is used for the sink terms.

The compressed version of $ \tens{B} $ is still very large
(several gigabytes for the
runs presented
in this work).
A disadvantage of the precalculation technique is therefore that the
size of the grid that can be used is restricted by the available amount of memory,
whereas a code that recalculates (parts of) $ \tens{B} $ at every time step
is only limited by CPU power.
A cubic dependence of the non-zero entries of $ \tens{B} $ on the number of mass bins
(as opposed to quadratic dependencies on the orbital element variables),
motivates the choice of a relatively small mass grid, representing only a part of
the collisional cascade.

\section{Model verification}
\label{s:app_verif}

We performed several tests to verify our numerical debris disk model.
The \mbox{P-R} drag and sublimation modules of the code were tested
by comparing the behavior of the model with
independent numerical solutions of the equations of motion and sublimation for individual particles.
For this purpose, the collisional module of the code was switched off,
and a single bin was filled as the initial setup of the model,
corresponding to the orbital elements of the particle.
For all these tests, the resulting evolution of the dust distribution (not shown here)
matches that of the independent solution.
The accuracy of the predictions is limited by numerical diffusion,
and becomes better with higher resolution (i.e., larger space-space grids).

To test the collisional module of our code, we benchmarked the model against the
code of \citet{2006A&A...455..509K}, by replicating one of their model runs for the debris disk of
Vega as accurately as possible.
Since the code of \citet{2006A&A...455..509K} uses semi-major axes
as the distance dimension of the phase-space grid,
the benchmark runs were performed with a version of our code that also uses the semi-major axis
(as opposed to periastron distance, used is the rest of this paper).
These runs do not include the effects of \mbox{P-R} drag or sublimation, so
they can be used to separately test the collisional module of our code, by switching
off \mbox{P-R} drag and sublimation.
Of the various runs presented by \citet{2006A&A...455..509K}, the specific one
that was reproduced is characterized by an initial optical depth profile
in the outer disk (beyond 120~AU)
that scales as $\tau\sub{geo}(r) \propto r^{-4}$, an initial eccentricity distribution between 0
and 0.375, and material properties for ``rocky'' grains.
We refer the reader to \citet{2006A&A...455..509K} for the specific values used for parameters describing
the phase-space grid, the initial setup of the disk, material properties, etcetera.

\begin{figure}[!t]
  \includegraphics[width=\linewidth]{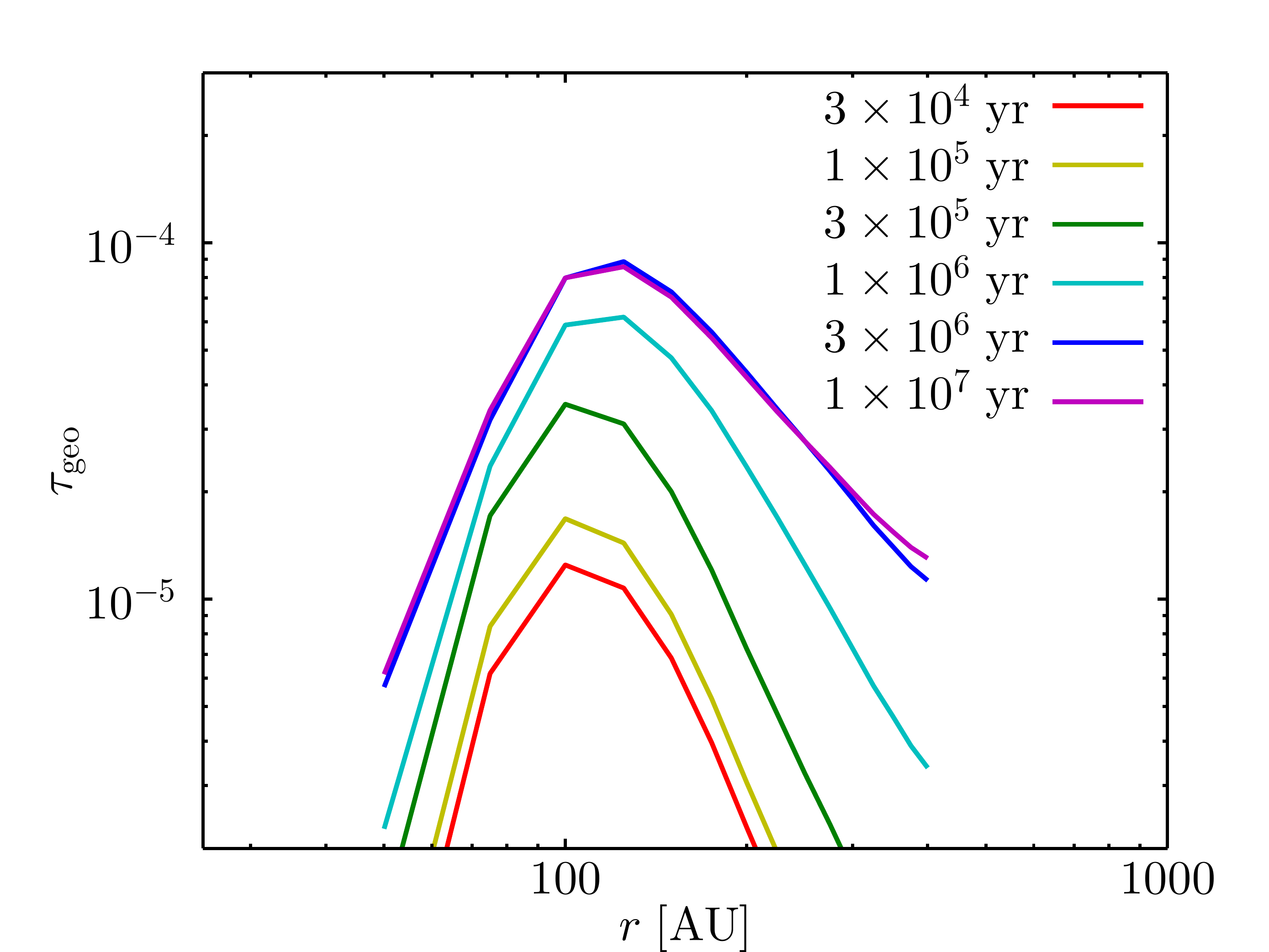}
  \caption{Evolution of the radial geometrical optical depth profile
  of the benchmark run, to be compared with Fig.~10 of \citet{2006A&A...455..509K}.}
  \label{fig:val_tau}
\end{figure}

\begin{figure}[!t]
  \includegraphics[width=\linewidth]{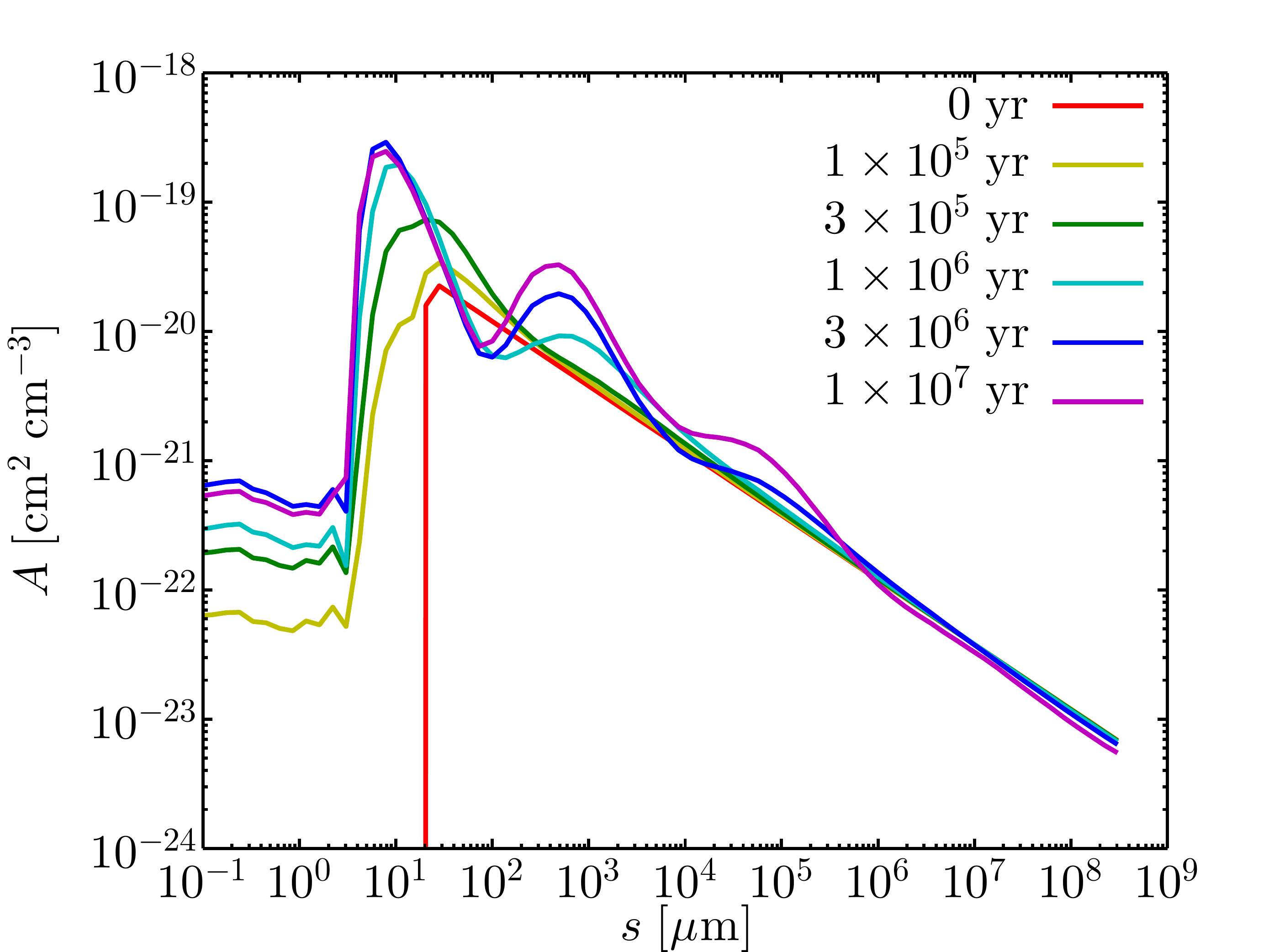}
  \caption{Evolution of the size distribution at $ r = 100 $~AU
  of the benchmark run, to be compared with Fig.~6 of \citet{2006A&A...455..509K}.}
  \label{fig:val_size}
\end{figure}

The evolution of the radial and size distribution
predicted by the benchmark run are shown
in Figs.~\ref{fig:val_tau} and \ref{fig:val_size}, respectively.
The corresponding results of \citet{2006A&A...455..509K} are their Figs.~10 and 6, respectively.
Comparison of the results reveals a good agreement between the two codes,
and the remaining discrepancies can be accounted for.
Relative to the benchmark, the our model predicts
(1) lower unbound particle populations,
and (2) shorter evolutionary timescales.
Point (1) is to be expected, since the unbound particle densities of \citet{2006A&A...455..509K} are
computed using the product of their production rate and their disk crossing time,
rather than from Eqs.~\ref{eq:unbound_lohne_basic} (T. L\"{o}hne, private communication).
We attribute point (2) to a mislabeling of the time steps in Figs.~6 and 10 of \citet{2006A&A...455..509K}.
Further discrepancies can be due to small differences in input parameters and numerical techniques used.

\section{Post-processing of model output}
\label{s:app_postproc}

The output of a model run are the population levels $ \vec{ n } $
of all bins in the \mbox{3-dimensional} phase space of particle mass and orbital elements,
as well as their change rates $ \vec{ \dot{ n } } $
(which should equal zero once a steady state is reached, except for bins corresponding to unbound orbits).
While this data is useful to analyze the orbital characteristics of different classes of particles in the debris disk,
it is often more convenient to know about the state of the disk as
a function of distance from the star.
This is essential, for example, if the results of the model are to be compared with observations.
Here, we describe the processing steps that are applied to the model output
to derive the quantities used in Sect.~\ref{s:results}.

\subsection{Conversion from orbital elements to radial distance}
\label{s:app_postproc_haug}

To find the radial distribution of matter in the debris disk,
the orbital element phase-space distribution function $n(m,q,e)$
(dimension: [g$^{-1}$~cm$^{-1}$])
needs to be converted to the configuration space distribution function $\mathcal{N}(m,r)$,
which denotes the
vertically-averaged number density of particles with masses
[$m,m+\dif m$] at distance $r$ (dimension: [g$^{-1}$~cm$^{-3}$]).
This problem was first solved by \citet{1958ZA.....44...71H}
for a rotationally symmetric ensemble of particles on Keplerian orbits.
Here, we give a brief derivation,
under the additional assumption that the distribution of inclinations is independent of
the distribution of the other orbital elements.

Consider an individual particle on a bound Keplerian orbit, that spends $ \dif t $ time
to cross a radial annulus with width $ \dif r $ at distance $r$ from the star.
The contribution of this particle to $\mathcal{N}(m,r)$ is
\begin{equation}
  \label{eq:haug_derive}
  \mathcal{N}\sub{part} =
  \frac{ 2 \dif t }{ P }
  \frac{ 1 }{ 2 \pi r \, \dif r }
  \frac{ 1 }{ h },
\end{equation}
where $P$ is the particle's orbital period,
and $h = 2 r \sin\varepsilon$ is the disk height, $\varepsilon$ being the semi-opening angle of the disk.
The explicit factor~2 in the numerator accounts for the fact that
the particle passes through this radial annulus twice during each orbit.
In terms of orbital elements $q$ and $e$, the orbital period $P$, accounting for direct radiation pressure, is given by
\begin{equation}
  \label{eq:period}
  P = 2 \pi \sqrt{ \frac{ q^3 }{ G M\sub{\star} ( 1 - \beta ) ( 1 - e )^3 } }.
\end{equation}
The radial velocity $ \dot r $ of the particle is given by
\begin{equation}
  \label{eq:rdot}
  \frac{ \dif r }{ \dif t }
  = \pm \sqrt{ \frac{ G M\sub{\star} ( 1 - \beta ) }{r}
    \left[ 2 - \frac{r}{q} (1 - e) - \frac{q}{r} (1 + e) \right] }.
\end{equation}
Combining Eqs.~\ref{eq:haug_derive}, \ref{eq:period}, and \ref{eq:rdot},
and considering all particles on bound orbits, gives \citep[cf.][]{2005Icar..174..105K}
\begin{equation}
  \begin{split}
    \label{eq:haug}
    \mathcal{N}(m,r) & = \frac{1}{ 4 \pi^{2} \sin (\varepsilon) } \frac{1}{ r^{3} }
      \int\limits_{q} \! \! \! \int\limits_{e} n(m,q,e) \left[ \frac{r}{q} (1-e) \right]^{3/2} \\
    & \quad \times \left[ 2 - \frac{r}{q}(1 - e) - \frac{q}{r}(1+e) \right]^{-1/2} \dif q \, \dif e,
  \end{split}
\end{equation}
with the integration domain
\begin{equation}
  \label{eq:haug_domain}
  \frac{1-e}{1+e} r \leq q \leq r, \qquad 0 \leq e < 1.
\end{equation}

The contribution of unbound particles to $\mathcal{N}(m,r)$ is calculated using their
production rate $\dot n(m,q,e)$ and the radial velocity with which they
leave the system (T. L\"{o}hne, private communication).
Assuming all unbound particles are created at the periastron of their orbits,
their vertically-averaged particle number density is given by
\begin{align}
  \label{eq:unbound_lohne_basic}
  \mathcal{N}(m,r) & = \frac{ 1 }{ 2 \pi r h } \int\limits_{q} \! \! \! \int\limits_{e}
    \frac{ \dot n(m,q,e) }{ | \dot r(m,q,e,r) | }
    \, \dif q \, \dif e \\
  \label{eq:unbound_lohne}
  \begin{split} & =
      \frac{ 1 }{ 4 \pi \sin\varepsilon \sqrt{ G M\sub{\star} ( 1 - \beta ) r^3 } }
    \int\limits_{q} \! \! \! \int\limits_{e} \dot n(m,q,e) \\
    & \quad \times \left[ 2 - \frac{r}{q} (1 - e) - \frac{q}{r} (1 + e) \right]^{-1/2}
    \dif q \, \dif e,
  \end{split}
\end{align}
with the integration domain
\begin{equation}
  \label{eq:unbound_lohne_domain}
  q \leq r, \qquad e \leq -1 \vee e \geq 1.
\end{equation}
Negative eccentricities correspond to ``anomalous'' hyperbolic orbits \cite[see][]{2006A&A...455..509K}.

In applying this theory to the raw data,
we replace the integrals in Eqs.~\ref{eq:haug} and \ref{eq:unbound_lohne} with sums;
replace $n(m,q,e)$ and $\dot n(m,q,e)$ with $ \vec{ n } $ and $ \vec{ \dot{ n } } $, respectively;
and evaluate the resulting equations at discrete points of $r$.
For each bin, we sample the phase-space it represents using a Monte Carlo method.

\subsection{Derived quantities}
\label{s:app_postproc_quantities}

For radial dust distribution profiles, we use
the vertical geometrical optical depth $\tau\sub{geo}$,
defined as
the surface density of collective (i.e., combining particles of all sizes) cross-section.
It is computed from $\mathcal{N}(m,r)$ as
\begin{align}
  \tau\sub{geo}(r)
  & = h \int \limits_{m} \sigma(m) \mathcal{N}(m,r) \, \dif m \\
  \label{eq:tau_eff}
  & = 2 \pi \sin(\varepsilon) r \int \limits_{m} s^{2}(m) \mathcal{N}(m,r) \, \dif m.
\end{align}

To characterize size distributions, we use the quantity $A(s,r)$,
defined as the cross-section density per \mbox{base-10} logarithmic unit of size.
It is given by
\begin{align}
  A(s,r)
    & = \frac{ \dif s }{ \dif \log\sub{10} (s) } \frac{ \dif m }{ \dif s } \sigma(s) \mathcal{N}(m,r) \\[0.5em]
    & = 4 \pi^{2} \ln(10) \rho\sub{d} s^{5} \mathcal{N}(m,r).
\end{align}
This quantity
allows for an easy comparison between
the relative contributions of particles with different sizes
to the total cross-section of the disk.
In the size distributions plots (Figs.~\ref{fig:size_prtail} and \ref{fig:size_pileup}),
a horizontal line means particles of all sizes contribute equally to the cross-section,
and equal areas under the curve correspond to equal contributions to the total cross-section.

\end{appendix}

\end{document}